\documentclass{article}
\usepackage{cite}
\usepackage{amsmath,amssymb,amsfonts}
\usepackage[noend]{algpseudocode}
\usepackage{algorithmicx,algorithm}
\usepackage{graphicx}
\usepackage{multirow}
\usepackage{subfigure}
\usepackage{setspace}
\usepackage{float}
\usepackage{color}
\usepackage[T1]{fontenc}
\usepackage[utf8]{inputenc}
\usepackage{authblk}
\newenvironment{proof}{{\noindent\it Proof}\quad}{\hfill $\square$\par}
\date{}

\title{\textbf{Robust Alignment of  Multi-Exposed Images with Saturated Regions}}
\author[1]{Jun Jiang}
\author[2]{Zhengguo Li} 
\author[2]{Shoulie Xie} 
\author[3]{Shiqian Wu} 
\author[1]{Liangcai Zeng \thanks{Corresponding author:zengliangcai@wust.edu.cn}}
\affil[1]{Key Laboratory of Metallurgical Equipment and Control Technology, Wuhan University of Science and Technology, Wuhan, China}
\affil[2]{Institute for Infocomm Research, Singapore}
\affil[3]{Institute of Robotics and Intelligent Systems, School of Information Science and Engineering, Wuhan University of Science and Technology, Wuhan, China}

\usepackage[a4paper,left=25mm,right=25mm,top=25mm,bottom=25mm]{geometry}  
\begin{document}
\maketitle
\begin{abstract}
It is challenging to align multi-exposed images due to large illumination variations, especially in presence of saturated regions. In this paper, a novel image alignment algorithm is proposed to cope with the multi-exposed images with saturated regions. Specifically, the multi-exposed images are first normalized by using intensity mapping functions (IMFs) in consideration of saturated pixels. Then, the normalized images are coded by using the local binary pattern (LBP). Finally, the coded images are aligned by formulating an optimization problem by using a differentiable Hamming distance. Experimental results show that the proposed algorithm outperforms state-of-the-art alignment methods for multi-exposed images in terms of alignment accuracy and robustness to exposure values.
\end{abstract}

\section{Introduction}
\label{sec:intro}
Image alignment is an important process in many image processing applications \cite{Szeliski2007Image}, visual odometry and SLAM \cite{1whel2015}, and panorama imaging \cite{1yaow2015}. A robust image alignment algorithm should cope with illumination variations and large motion variations while giving a sub-pixel accurate alignment. Existing image alignment approaches can be classified into two categories \cite{Szeliski2007Image}: intensity-based methods and feature-based methods. The intensity-based methods find motion parameters directly from image intensities on brightness constancy assumption. In contrast, the feature-based methods, working with image features (such as SIFT-based features), are usually used to deal with illumination-varying images.

Since the brightness constancy assumption is not true for differently exposed images \cite{debevec2008recovering}, the feature-based methods are widely applied to align differently exposed images. Ordering features are usually used to represent invariant property of illumination-variation images. Median threshold bitmap (MTB) \cite{ward2003fast} is a global descriptor for the representation of  illumination-variation images. But the alignment is not accurate due to errors from unequal division on severely saturated regions and ordering problems for many pixels with the same intensity \cite{1wu2014}. Local descriptors, such as census transform (CT) \cite{1zabih1994}, local binary pattern (LBP) \cite{ojala1996comparative}, BRIEF \cite{calonder2011brief}, ORB \cite{Rublee2012ORB} and so on, have been widely used to describe illumination-robust features. These descriptors depend solely on comparison between a specific pixel and its neighborhood (or a pair of pixels in its neighborhood), and are therefore invariant under intensity changes. 

Several learning-based binary descriptors have been proposed to project a local patch to a binary descriptor.  Semantic hashing \cite{2009Semantichashing} trains a multi-layer neural network to learn compact binary codes. Spectral hashing (SpeH) \cite{SpectralHashing} generates efficient binary codes by spectral graph partitioning. BinBoost descriptor \cite{BinBoost} applies boosting to learn a set of binary hash functions that achieve an accurate binary descriptor. The work in \cite{2015BOLD} presented a binary online learned descriptor by applying LDA criterion. DeepBit \cite{DeepBit} generates a compact binary descriptor by designing a CNN in an unsupervised manner. The work in \cite{LearningCBFD} proposed a compact binary face descriptor (CBFD) to learn evenly-distributive and energy-saving local binary codes. Duan et al. \cite{Duan2017CALBP} presented a context-aware local binary feature learning (CA-LBFL) approach using the contextual information of adjacent bits. A supervised structured binary code (SUBIC) with a one-hot block structure was proposed in \cite{Jain2017SUBIC}. Those methods utilize the rigid sign function or a hand-crafted threshold for binarization, which is easily affected by exposure change and leads to unequal binarization \cite{Duan2019}. Normally, the learning-based matching pipeline consists of three sub-components \cite{He2018LocalDescriptors}: keypoints detection, orientation estimation and features extraction. Each subcomponent is trained individually. However, the overall performance may not be optimal when integrating these separately optimized subcomponents into a matching pipeline \cite{2016LIFT}. Therefore, end-to-end trainable matching networks \cite{2017Superpoint,2018LFnet,2019RFnet} have been proposed to optimize the matching performance. But, they require hand-crafted detectors or other functions to initialize the training process \cite{2018LFnet,2019RFnet}. More importantly, the learning-based methods are very data-dependent and require a large amount of data to train the network repeatedly, which is complex, time-consuming and not suitable for using on mobile devices. In contrast, the traditional binary descriptors is a better and simpler choice to represent the differently exposed images.

For binary descriptors \cite{calonder2011brief, Rublee2012ORB, 2015BOLD}, Hamming distance was adopted to measure the similarity of two pixels, and can be computed with a bitwise XOR operation followed by a bit count. However, the Hamming distance is non-differentiable and unsuitable for optimization. Wu et al. \cite{1wu2014} used the decimal representation to represent the LBP coded images and matched the difference between two coded images under the squared Euclidean norm.  However, the decimal representation does not capture the closeness of two bit-strings and it is sensitive to the rotation. It is thus desired to have a differentiable distance which is equivalent to the Hamming distance  for aligning of differently exposed images. Such a distance function was recently introduced in \cite{1alis2017}. Besides the issues on the brightness constancy assumption and the distance, one more challenging issue for aligning multi-exposed images is possibly saturated regions in the images \cite{1alis2017}. Due to the saturated regions, one feature detected in one image may not occur on another; and a specific intensity in one image may map to multiple intensities in the other image, and vice versa.  It is thus desired to develop a new image alignment algorithm for multi-exposed images with saturated regions.

In this paper, a novel alignment algorithm is proposed for multi-exposed images with saturated regions by using the differentiable ``Hamming'' distance function to binary descriptors in \cite{1alis2017}. Specifically, a novel intensity mapping function (IMF) based method is first proposed to normalize the multi-exposed images so that the saturated regions are same when they are aligned. Then the normalized images are coded by the LBP. The difference between two coded images is measured by the differentiable ``Hamming'' distance in \cite{1alis2017} which is equivalent to the Hamming distance on top of the bitwise XOR. Same as \cite{1wu2014}, the alignment of two differently exposed images is formulated as a quadratic optimization which can be easily solved. Since the differentiable ``Hamming'' distance captures the closeness of two bit-strings better than the decimal representation in \cite{1wu2014}, the proposed algorithm is more robust than the algorithm in \cite{1wu2014}.  Experimental results demonstrate that the proposed algorithm outperforms state-of-the-art alignment algorithms for multi-exposed images. Two major contributions of this paper are summarized as follows: 1)  A new normalization method for differently exposed images is proposed which is effective to reduce the effect of saturation to image alignment; and 2) A simple image alignment algorithm for differently exposed images is presented which is robust to the exposure values (EV).

The rest of this paper is organized as follows. Multi-exposed images are normalized in section \ref{section1} and aligned in section \ref{section2}. Experimental results are provided in section \ref{section3} to validate the proposed algorithm. Finally, conclusions are provided in section \ref{section4}.

\section{Normalization of  Multi-Exposed Images}
\label{section1}

In this section, a novel normalization method is first introduced for  multi-exposed images. Unlike the method in \cite{1engel2018}, the exposure times of the images are not required by the proposed method.

Consider two images $Z_1$ and $Z_2$ of an identical scene with two
different exposures $\Delta t_1$ and $\Delta t_2$. $Z_1(p)(Z_2(p))$, $Z_1(p')(Z_2(p'))$ and $E(p)(E(p'))$
are two intensities and radiances respectively at pixel positions $p$ and $p'$. There are two possible issues for two differently exposed images. One is that their intensity values are different for two co-located pixels. Fortunately, the following relationship is generally held due to the monotonic
property:
\begin{equation}
E(p)>E(p') \Rightarrow Z_1(p)>Z_1(p'), \mbox{~and~} Z_2(p)>Z_2(p').
\end{equation}

Clearly, exposures change the intensities, but keep the relative order of intensities if they are not saturated. Thus, the LBP code could be adopted to represent two differently exposed images \cite{1wu2014}.

One possible issue for multi-exposed images is the saturation, i.e.,  part of an image could be over/under-exposed. The saturation is a very challenging issue for alignment \cite{1alis2017}. Assume that $\Delta t_1$ is larger than $\Delta t_2$. Let $\alpha$  and  $\beta$  be two predefined constants, and their values are selected as 5 and 254, respectively. For simplicity, a pixel $Z(p)$ is regarded as over-exposed if its value is larger than $\beta$, and  under-exposed if its value is smaller than $\alpha$. Here, a novel method is introduced to synchronize the under-exposed regions of the images $\hat{Z}_1$ and $\hat{Z}_2$ as well as the over-exposed regions of the images $\hat{Z}_1$ and $\hat{Z}_2$ by using the IMFs between the images $Z_1$ and $Z_2$.

Let $f_{12}(z)$  and $f_{21}(z)$ be the IMFs from image $Z_1$  to image $Z_2$    and vice versa, respectively. By finding the matched points between the cumulative histograms of $Z_1$ and $Z_2$, the IMF $f_{12}(z)$ which maps intensity from $Z_1$ to $Z_2$ and the inverse IMF $f_{21}(z)$ can be obtained \cite{grossberg2003determining}.

\begin{figure*}
	\centering
	{
		\includegraphics[width=0.8\textwidth]{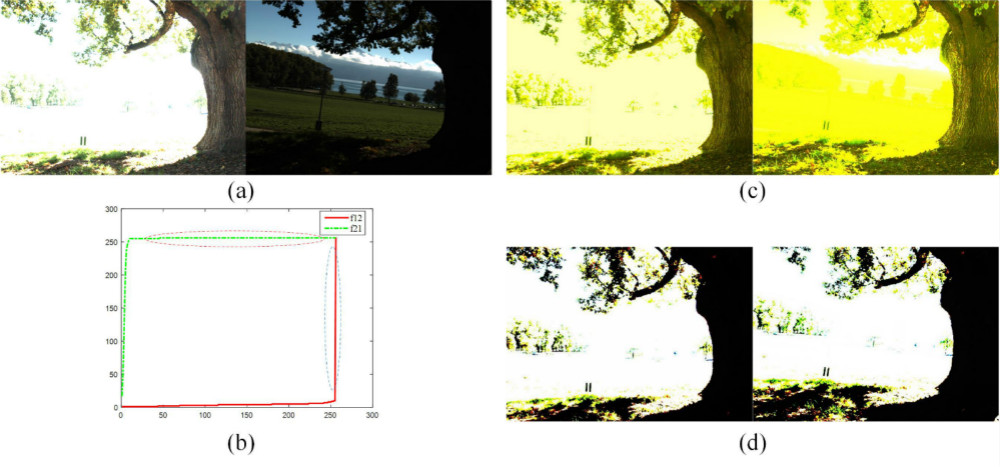} 		
	}	
	\caption{ Normalization of 1st and 6th images in ``BigTree''. (a) Original images. The right image is rotated by 5$ ^{\circ} $, and shifted by 30 pixels and 10 pixels in y-axis and x-axis respectively. (b) IMFs of the two images. (c) Normalized images using a unidirectional mapping function\cite{1wu2014}. It maps the image with more information to the one with less information. (d) Normalized images using the proposed bi-directional mapping function. The intensities larger than $\zeta_1$ in the left image are unchanged, the remaining intensities in right image are mapped to left image to ensure the over-exposure regions in both images are consistent. The intensities less than $\zeta_2$ in the right image remain constant, the remaining intensities in the left image are mapped to right image to ensure the under-exposure regions in both images are consistent.} 
	\label{Fig4}
\end{figure*}

It was pointed out in \cite{18li2014} that the IMFs $f_{12}(z)$ and $f_{21}(z)$ have the following properties: 1) The IMF $f_{12}(Z_1(p))$ is not accurate if $Z_1(p)$ is over-exposed; and 2) The IMF $f_{21}(Z_2(p))$ is not accurate if $Z_2(p)$ is under-exposed.
On top of the two observations, two constants are defined as

\begin{equation}
\left\{\begin{array}{l}
\label{equ2}
\zeta_1=\max_{z_{1}}\{z_1|f_{12}(z_{1})=\alpha\}\\
\zeta_2=\min_{z_{2}}\{z_2|f_{21}(z_{2})=\beta\}
\end{array}
\right.,
\end{equation}

$\zeta_1$ and $\zeta_2$ are further adjusted as
\begin{align}
\left\{\begin{array}{l}
\zeta_1=\min\{\zeta_1,\beta\}\\
\zeta_2=\max\{\zeta_2,\alpha\}
\end{array}
\right..
\end{align}
The image  $Z_1$ is  mapped as
\begin{equation}
\label{equ3}
\hat{Z}_1(p)=\left\{\begin{array}{ll}
Z_1(p); &\mbox{if~}Z_1(p)\geq \zeta_1\\
f_{12}(Z_1(p)); &\mbox{otherwise}
\end{array}
\right.,
\end{equation}
and the image $Z_2$  is mapped as
\begin{equation}
\label{equ3}
\hat{Z}_2(p)=\left\{\begin{array}{ll}
Z_2(p); &\mbox{if~}Z_2(p)\leq \zeta_2\\
f_{21}(Z_2(p)); &\mbox{otherwise}
\end{array}
\right..
\end{equation}
It can be easily shown that
\begin{align}
\label{equ5}
\left\{\begin{array}{l}
\hat{Z}_1(p)\in [0,  \alpha] \cup [\zeta_1, 255]\\
\hat{Z}_2(p)\in [0,  \zeta_2]\cup [\beta, 255]
\end{array}
\right..
\end{align}
Clearly, the under-exposed regions of $\hat{Z}_1$ and $\hat{Z}_2$ as well as the over-exposed regions of $\hat{Z}_1$ and $\hat{Z}_2$ are the same if they are aligned.
The translational vector and the rotational angle are estimated by using the mapped images $\hat{Z}_1$  and $\hat{Z}_2$.

As the functions $f_{12}(\cdot)$ and  $f_{21}(\cdot)$ are monotonically non-decreasing functions, and
\begin{equation}
\left\{\begin{array}{l}
f_{12}(Z_1(p))\leq Z_1(p)\\
f_{21}(Z_2(p))\geq Z_2(p)
\end{array}
\right..
\end{equation}
it can be easily verified that the LBP codes of the mapped images $\hat{Z}_1$ and  $\hat{Z}_2$ are almost the same as those of the original images $Z_1$ and $Z_2$. Clearly, both the IMF $f_{12}(\cdot)$  and the IMF $f_{21}(\cdot)$ are applied to improve the robustness of LBP codes with respect to possible saturation in the proposed algorithm. The IMFs of images 1 and 6 in ``BigTree'' sequence is plotted in Fig.\ref{Fig4}(a). As shown in Fig.\ref{Fig4}(c), the IMFs result in inconsistence in the two normalized images, such as trees in bright area, and tree trunk in dark area. This is because the region circled by the two ellipses in Fig.\ref{Fig4}(b) is not reliable due to under/over exposure and multi-value mapping. Thus, the parameters  $ \alpha $  and $ \beta $ are introduced to reduce their effects, and improve the similarity of the two normalized images, as shown in Fig.\ref{Fig4}(d).

\section{Alignment of Multi-Exposed Images}
\label{section2}

In this section, a new alignment algorithm is proposed for multi-exposed images by using a differentiable ``Hamming'' distance \cite{1alis2017}.

Let the luminance components of the two normalized images $\hat{Z}_1$ and $\hat{Z}_2$ be  denoted as $Y_1$ and $Y_2$, respectively. In order to reduce the sensitivity of the descriptor to noise, the images $Y_1$ and $Y_2$ can be smoothed with a Gaussian filter in a $3\times 3$ neighborhood ($\sigma=0.5$) \cite{1alis2017} or an edge-preserving smoothing filter such as the weighted guided image filter (WGIF) in a $3\times 3$ neighborhood ($\lambda=1/256$) \cite{2li2014}. For the pixel $p$, $\chi(j)(j=1,2,\cdots,8)$ are its eight neighboring pixels. The Census Transform or the LBP code is  \cite{1zabih1994, ojala1996comparative}
\begin{equation}
\label{equ7}
S^{(j)}_{Y_i}(p)=\left\{\begin{array}{ll}
1; &\mbox{if~}Y_i(\chi(j))\bowtie Y_i(p)\\
0; &\mbox{otherwise}\\
\end{array}
\right.,
\end{equation}
where the operation $\bowtie\in \{>, <, \geq, \leq\}$. Since the binary  descriptor requires only eight comparisons, it is commonly stored as a byte
according to \cite{1wu2014}
\begin{equation}
\label{distance123}
S_{Y_i}(p)=\sum_{j=1}^82^{j-1}S^{(j)}_{Y_i}(p).
\end{equation}

Unfortunately, the above binary  descriptor does not maintain the morphological invariance to intensity changes due to different weights for different neighborhood pixels. For example, consider two pairs of bit-strings   $a = \{1,0,1,0,1,1,1,0\} $ and  $b = \{1,1,1,0,1,1,1,0\} $, $c = \{1,1,1,0,1,1,1,1\} $ and  $d = \{1,1,1,0,1,1,1,0\} $. Both the pairs differ at a single bit, and they are clearly similar.  However, if the equation (\ref{distance123}) is used and matched under Euclidean norm, their distance become $64$ and $1$, respectively. Clearly, the equation (\ref{distance123}) does not capture the closeness in the descriptor space and it is sensitive to the rotational. To address this problem, the binary  descriptor must be matched using a binary norm, such as the Hamming distance, which counts the number of mismatched bits as follows:
\begin{equation}
\label{costj}
\rho_h(S_{Y_i}(p), S_{Y_i}(p'))=\sum_{j=1}^8(S^{(j)}_{Y_i}(p) \odot S^{(j)}_{Y_i}(p')).
\end{equation}
where $\rho_h(a,b)$ is Hamming distance denoted as $a\odot b$ and $\odot$ is the XOR operation. It is noted that equation (\ref{costj}) cannot be formulated as an optimization problem due to the non-differentiability of the Hamming distance. As such, an equivalently differentiable measurement is derived in the following.

{\it Proposition 1:} The Hamming distance in the equation (\ref{costj}) is equivalent to the following differentiable cost distance:
\begin{equation}
\label{costj1}
\rho(S_{Y_i}(p), S_{Y_i}(p'))=\sum_{j=1}^8(S^{(j)}_{Y_i}(p)-S^{(j)}_{Y_i}(p'))^2.
\end{equation}
\begin{proof}
	Notice that
	\begin{align*}
	a\odot b&=\left\{\begin{array}{ll}
	0;&\mbox{if~}\{a=0, b=0\} \mbox{~or~}\{a=1, b=1\}\\
	1; & \mbox{if~}\{a=1, b=0\} \mbox{~or~}\{a=0, b=1\}\\
	\end{array}
	\right.,\\
	(a-b)^2&=\left\{\begin{array}{ll}
	0;&\mbox{if~}\{a=0, b=0\} \mbox{~or~}\{a=1, b=1\}\\
	1; & \mbox{if~}\{a=1, b=0\} \mbox{~or~}\{a=0, b=1\}\\
	\end{array}
	\right..
	\end{align*}
	It can be easily shown that
	\begin{align}
	(a-b)^2=a\odot b
	\end{align} holds if both $a$ and $b$ are in the set $\{0,1\}$. Therefore, the Hamming distance in the equation (\ref{costj}) is equivalent to the differentiable cost distance in the equation (\ref{costj1}).
\end{proof}

The difference of two LBP coded images is computed as
\begin{equation}
\label{costjjj}
J=\sum_{p}\rho(S_{Y_1}(p), S_{Y_2}(\psi(p))).
\end{equation}
where $\psi(p)$ is the corresponding pixel in the image $\hat{Z}_2$ .

Let $p$ be denoted as $(x,y)$. Same as \cite{1wu2014}, it is assumed that the motion between the images $\hat{Z}_1$ and $\hat{Z}_2$ is the Euclidean transformation, i.e.,
\begin{equation}
\label{equ14}
\psi(p) = [x, y] R^T(\theta)+[t_x, t_y],
\end{equation}
where $R(\theta)$ is a 2D rotational matrix.  The values of $\theta$, $t_x$, and $t_y$ are obtained from the following optimization problem:
\begin{equation}
\label{optimizationproblem123}
\arg\min_{\theta,t_x, t_y}J(\theta,t_x, t_y).
\end{equation}

If the value of $\theta$, $t_x$ and $t_y$ are small, then image $ S_{Y_2} $ can be derived as below \cite{1wu2014},

\begin{eqnarray*}
	\label{lwu2014}
	\nonumber
	&&\hspace{-7mm}S_{Y_2}^{(j)}(\psi(p))\\
	&&\hspace{-7mm}=S_{Y_1}^{(j)}([x, y] R^T(\theta)+[t_x, t_y])\\ \nonumber
	&&\hspace{-7mm}\approx S_{Y_1}^{(j)}(x-y\theta+t_x, y+x\theta+t_y)\\\nonumber
	&&\hspace{-7mm}\approx S_{Y_1}^{(j)}(x,y)+\frac{\partial S_{Y_1}^{(j)}}{\partial x}(t_x-y\theta)+\frac{\partial S_{Y_1}^{(j)}}{\partial y}(x\theta+t_y)\\
	&&\hspace{-7mm}=S_{Y_1}^{(j)}(x,y)+\frac{\partial S_{Y_1}^{(j)}}{\partial x}t_x+\frac{\partial S_{Y_1}^{(j)}}{\partial y}t_y+(\frac{\partial S_{Y_1}^{(j)}}{\partial y}x-\frac{\partial S_{Y_1}^{(j)}}{\partial x}y)\theta.
\end{eqnarray*}

For simplicity, define a $3\times 3$ symmetric matrix $A$ as
\begin{eqnarray}
\left\{\begin{array}{l}A_{11} = \sum_{x,y}\sum_{j=1}^8(\frac{\partial S_{Y_1}^{(j)}}{\partial x})^2\\
A_{12} = A_{21}=\sum_{x,y}\sum_{j=1}^8\frac{\partial S_{Y_1}^{(j)}}{\partial y}\frac{\partial S_{Y_1}^{(j)}}{\partial x}\\
A_{13} = A_{31}=\sum_{x,y}\sum_{j=1}^8(\frac{\partial S_{Y_1}^{(j)}}{\partial y}x-\frac{\partial S_{Y_1}^{(j)}}{\partial x}y)\frac{\partial S_{Y_1}^{(j)}}{\partial x}\\
A_{22} = \sum_{x,y}\sum_{j=1}^8(\frac{\partial S_{Y_1}^{(j)}}\partial{ y})^2\\
A_{23} = A_{32}=\sum_{x,y}\sum_{j=1}^8(\frac{\partial S_{Y_1}^{(j)}}{\partial y}x-\frac{\partial S_{Y_1}^{(j)}}{\partial x}y)\frac{\partial S_{Y_1}^{(j)}}{\partial y}\\
A_{33}=\sum_{x,y}\sum_{j=1}^8(\frac{\partial S_{Y_1}^{(j)}}{\partial y}x-\frac{\partial S_{Y_1}^{(j)}}{\partial x}y)^2\\
\end{array}
\right.,
\end{eqnarray}
and a vector $b=[b_{1},b_{2},b_{3}]^T$ as
\begin{eqnarray}
\left\{\begin{array}{l}
b_1 = \sum_{x,y}\sum_{j=1}^8\frac{\partial S_{Y_1}^{(j)}}{\partial x}(S^{(j)}_1-S_{Y_2}^{(j)})\\
b_2 = \sum_{x,y}\sum_{j=1}^8\frac{\partial S_{Y_1}^{(j)}}{\partial y}(S^{(j)}_1-S_{Y_2}^{(j)})\\
b_2 = \sum_{x,y}\sum_{j=1}^8(\frac{\partial S_{Y_1}^{(j)}}{\partial y}x-\frac{\partial S_{Y_1}^{(j)}}{\partial x}y)(S^{(j)}_1-S_{Y_2}^{(j)})\\
\end{array}
\right..
\end{eqnarray}
It is easily shown that the equation $A\left[\begin{array}{lll}\theta &t_x & t_y\end{array}
\right]^{T}=b$ can be derived from the optimization problem (\ref{optimizationproblem123}).
Once the optimal values of $\theta^*$, $t_x^*$, and $t_y^*$ are obtained, the image $Z_2$ is aligned as
$Z_2([x, y] R^T(-\theta^*)+[-t_x^*, -t_y^*])$.

The optimization is very fast due to linear regression, and the solution can be achieved to sub-pixel accuracy. It is noted that the Taylor series involved in the estimation of $(\theta,t_x,t_y)$ is a good approximation only when the motion is very small. To further increase speed and robust to large motion, the optimization is implemented by coarse-to-fine technique which is represented using the Gaussian pyramid \cite{1wu2014}, shown in Algorithm \ref{111}, where the wrap() function is an affine transformation and is used to update the position of pixels.

\begin{algorithm}[t]
	\begin{spacing}{1.25}		
		\caption{Pseudo code of the proposed method} 
		\label{111}
		\hspace*{0.02in} {\bf Input:} 
		Two differently exposed images $Z_1$(Refernce images) and $Z_2$(Slave image). \\
		\hspace*{0.02in} {\bf Output:} 
		Motion parameters between image pairs $Z_{1}$ and $Z_{2}$	    
		\begin{algorithmic}[1]
			\State $\hat{Z}_1$,$\hat{Z}_2$ = $Normalization(Z_1,Z_2) $ 
			
			\State Estimate $t_x^{0}$, $t_y^{0}$ and $\theta^{0}$ by historgram-based matching
			
			\State $\hat{Z}_1$ = Wrap $(\hat{Z}_1,t_x^{0},t_y^{0},\theta^{0})$
			
			\State Build N-tier pyramid Gaussian pyramid of $\hat{Z}_1$,$\hat{Z}_2$
			
			\For{\textit{pyrlevel} = N:1} 
			
			\State Code $\hat{Z}_1^{pyrlevel}$, $\hat{Z}_2^{pyrlevel}$ by equation (\ref{equ7})
			
			\State Estimate $t_x^{pyrlevel}$, $t_y^{pyrlevel}$ and $\theta^{pyrlevel}$ by  ``Hamming Distance''	
			
			\State Update $t_x^{pyrlevel-1}=2*t_x^{pyrlevel}$, 
			$t_y^{pyrlevel-1}=2*t_y^{pyrlevel}$, 
			$\theta^{pyrlevel-1}=\theta^{pyrlevel}$
			
			\State Wrap $\hat{Z}_1^{pyrlevel-1}$ with $t_x^{pyrlevel}$, $t_y^{pyrlevel}$ , $\theta^{pyrlevel}$
			
			\EndFor		
			
			\State \Return $t_x^*$, $t_y^*$ and $\theta^*$
		\end{algorithmic}
	\end{spacing}
\end{algorithm}

It should be pointed out that the single scale with a few iterations is enough if the initial values of $\theta$, $t_x$ and $t_y$ can be computed by using the inertial measurement unit (IMU) integration \cite{John2018Motion} or other methods, such as histogram-based matching \cite{1wu2014}. Besides, the LBP code is also a simple way to reduce the effects of haze \cite{2018Singleimagehaze}.

\section{Experimental Results}
\label{section3}

In this section, the proposed method is evaluated with a variety of synthesized images from benchmark datasets, public datasets (Cai's dataset) \cite{2018DeepSingleImageContrastEnhancer}  and real images. All datasets are released in open source\cite{data}. Readers can browse the full-size images in the digital version and zoom in to better observe the differences. The results of the proposed method are compared with: 1) existing non-parametric ordering features using MTB \cite{ward2003fast}, CT \cite{1zabih1994}, LBP \cite{ojala1996comparative} and BRIEF \cite{calonder2011brief} methods, while employing the alignment algorithm shown in Section \ref{section2}; 2) the intensity-based method IMF; 3) the feature-based method SIFT \cite{Lowe2004Distinctive} and the hybrid method IMF+SIFT; 4) IMF+BRIEF and IMF+BRIEF+``Hamming'' distance (``HD''); 5) IMF+LBP\cite{1wu2014}; and 6) Learning-based matching methods SuperPoint \cite{2017Superpoint} and LF-Net \cite{2018LFnet}.

\subsection{Test on synthetic images}
\begin{figure*}
	\centering	
	\vspace{-0.35cm}
	\subfigure[``BigTree'' sequence. Fields, trees and mountains in first image are over-exposed.]{	
			\includegraphics[width=0.14\linewidth]{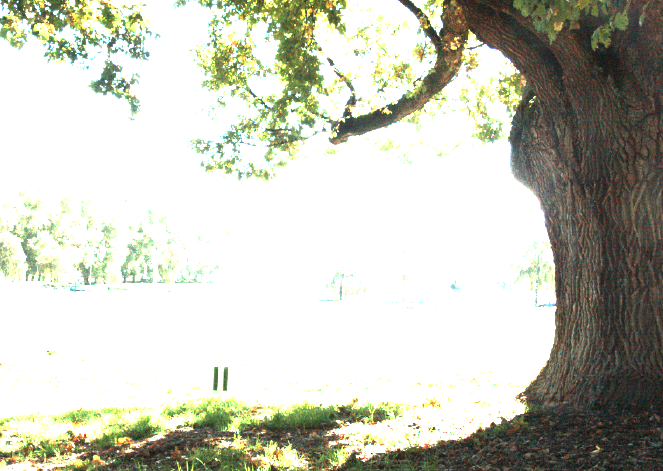}
			\includegraphics[width=0.14\linewidth]{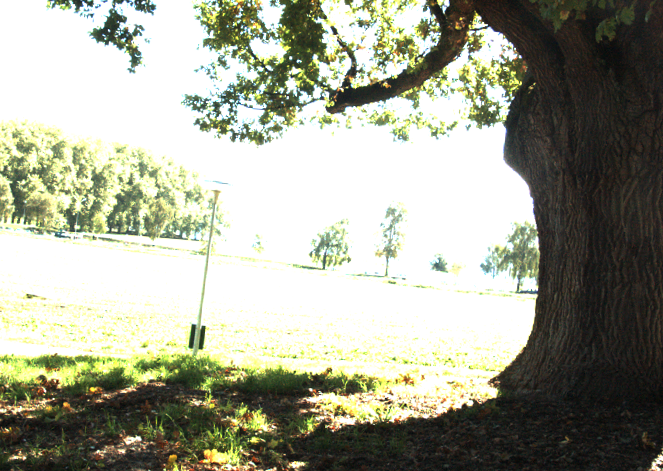}
			\includegraphics[width=0.14\linewidth]{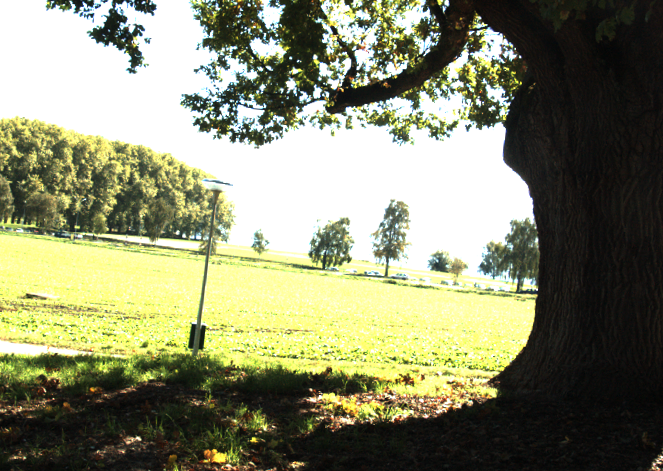}
			\includegraphics[width=0.14\linewidth]{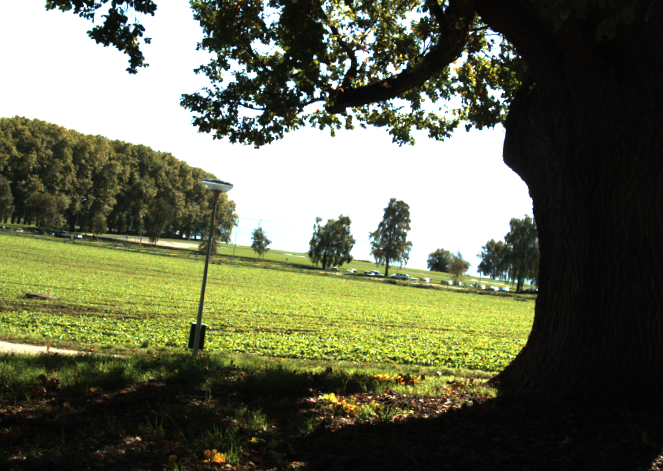}
			\includegraphics[width=0.14\linewidth]{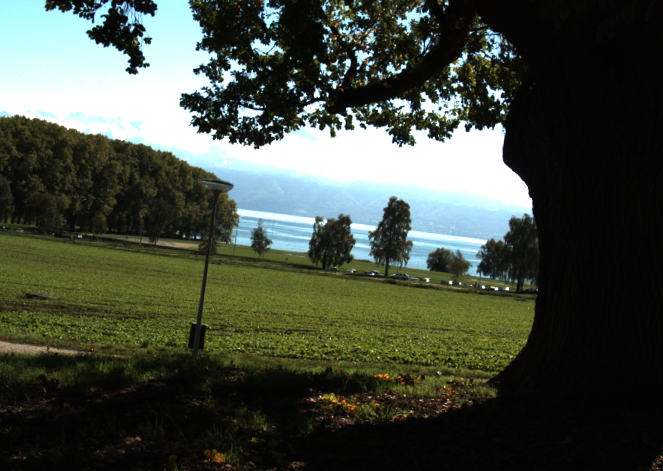}
			\includegraphics[width=0.14\linewidth]{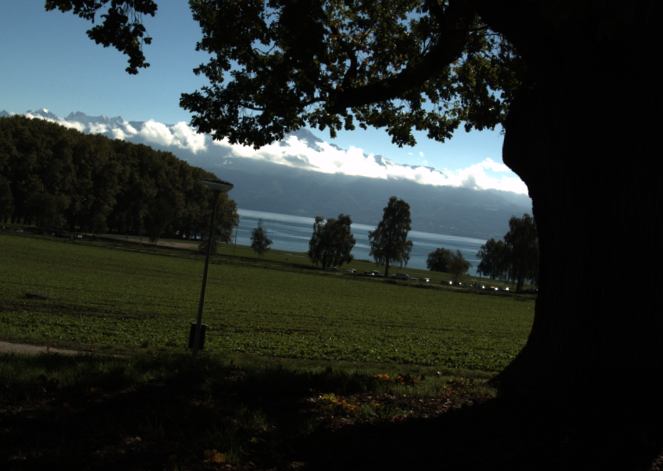}
			\includegraphics[width=0.14\linewidth]{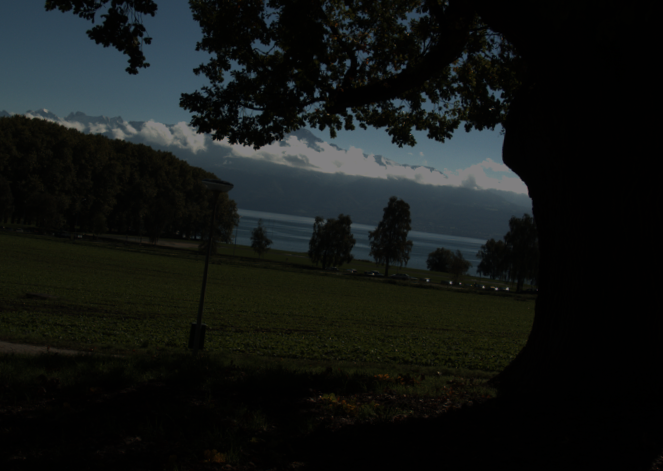}}
		
	\subfigure[``Snowman'' sequence. The first image is almost over-exposed due to sun glare.]{
		\includegraphics[width=0.165\linewidth]{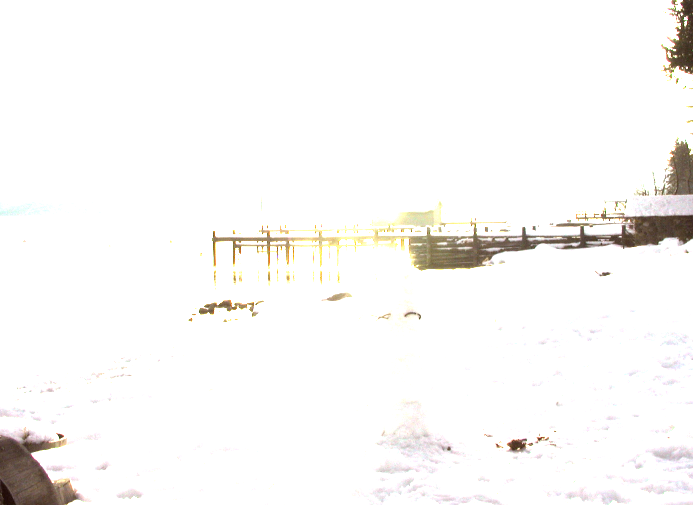}
		\includegraphics[width=0.165\linewidth]{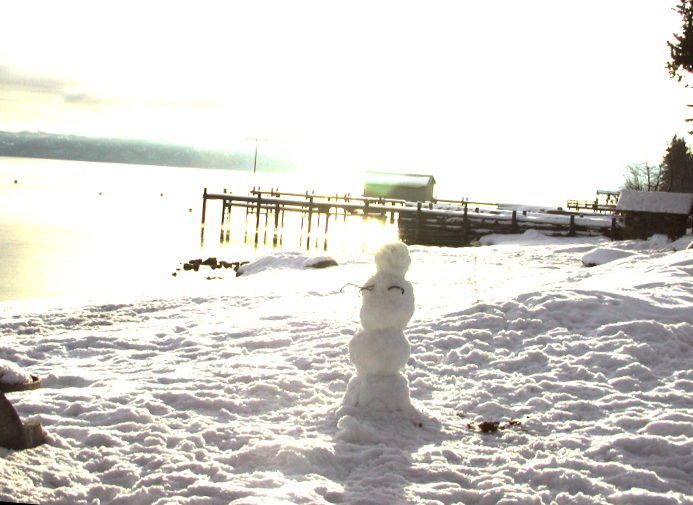}
		\includegraphics[width=0.165\linewidth]{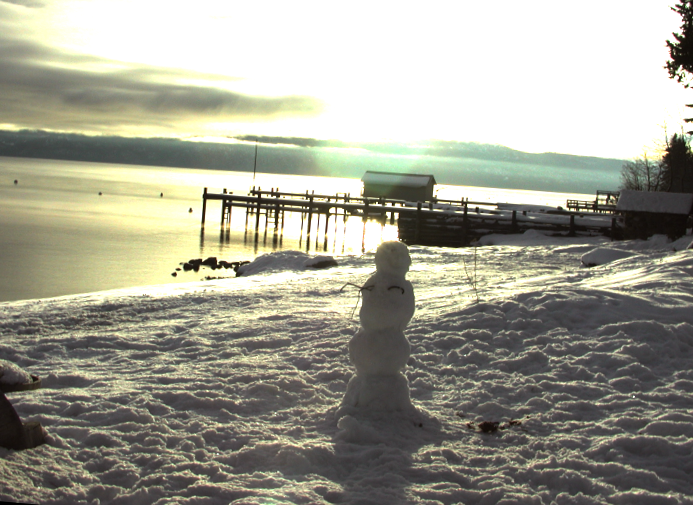}
		\includegraphics[width=0.165\linewidth]{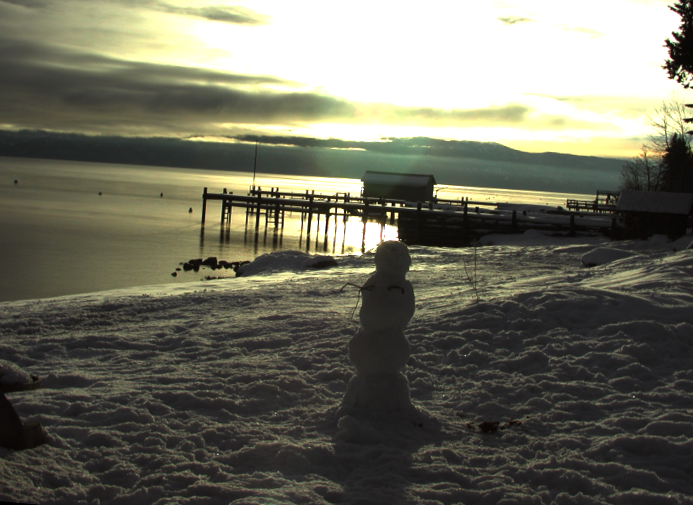}
		\includegraphics[width=0.165\linewidth]{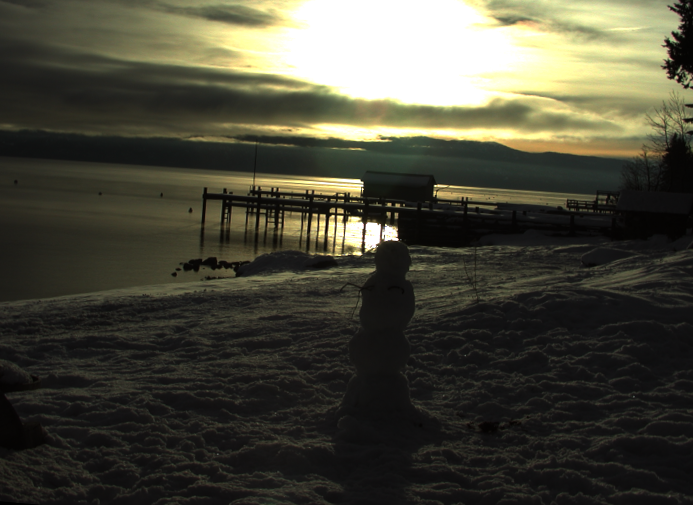}
		\includegraphics[width=0.165\linewidth]{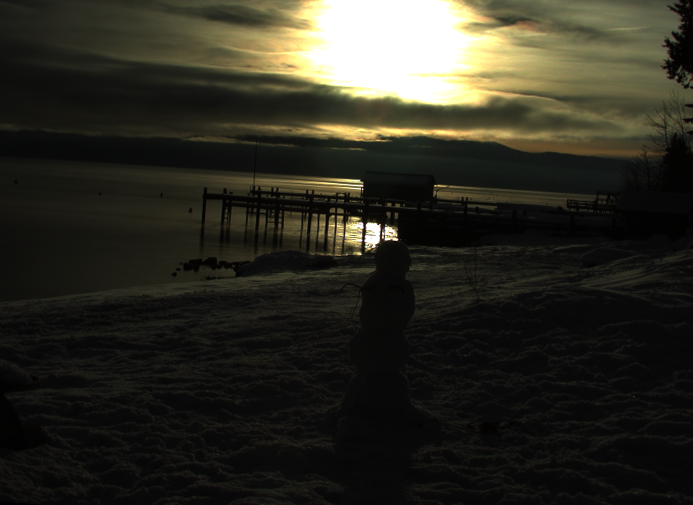}}
	
	\subfigure[``Inscription'' sequence. The text on the inscription in the first image is over-exposed and cannot be seen clearly.]{		
		\includegraphics[width=0.108\linewidth]{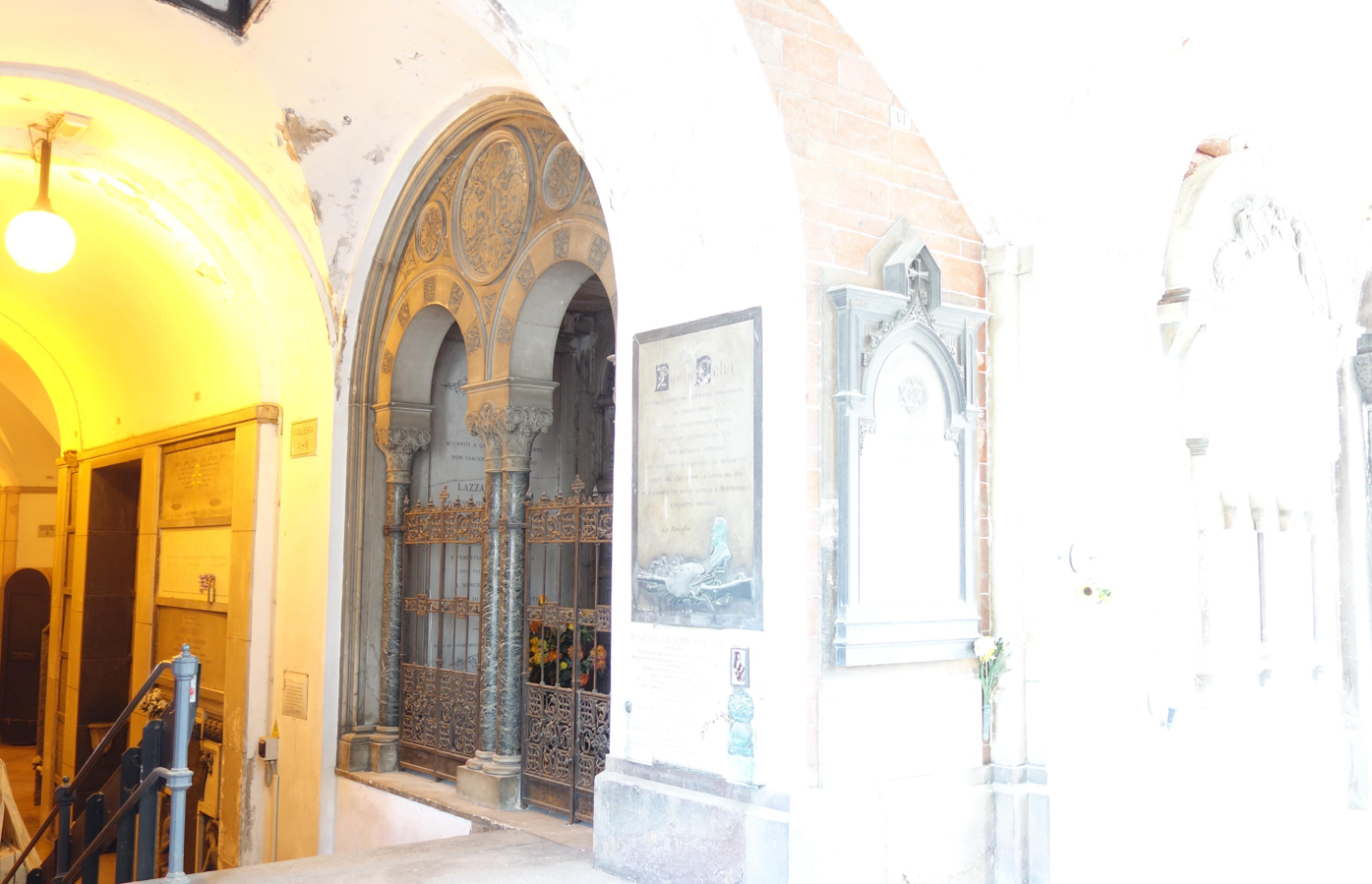} 		
		\includegraphics[width=0.108\linewidth]{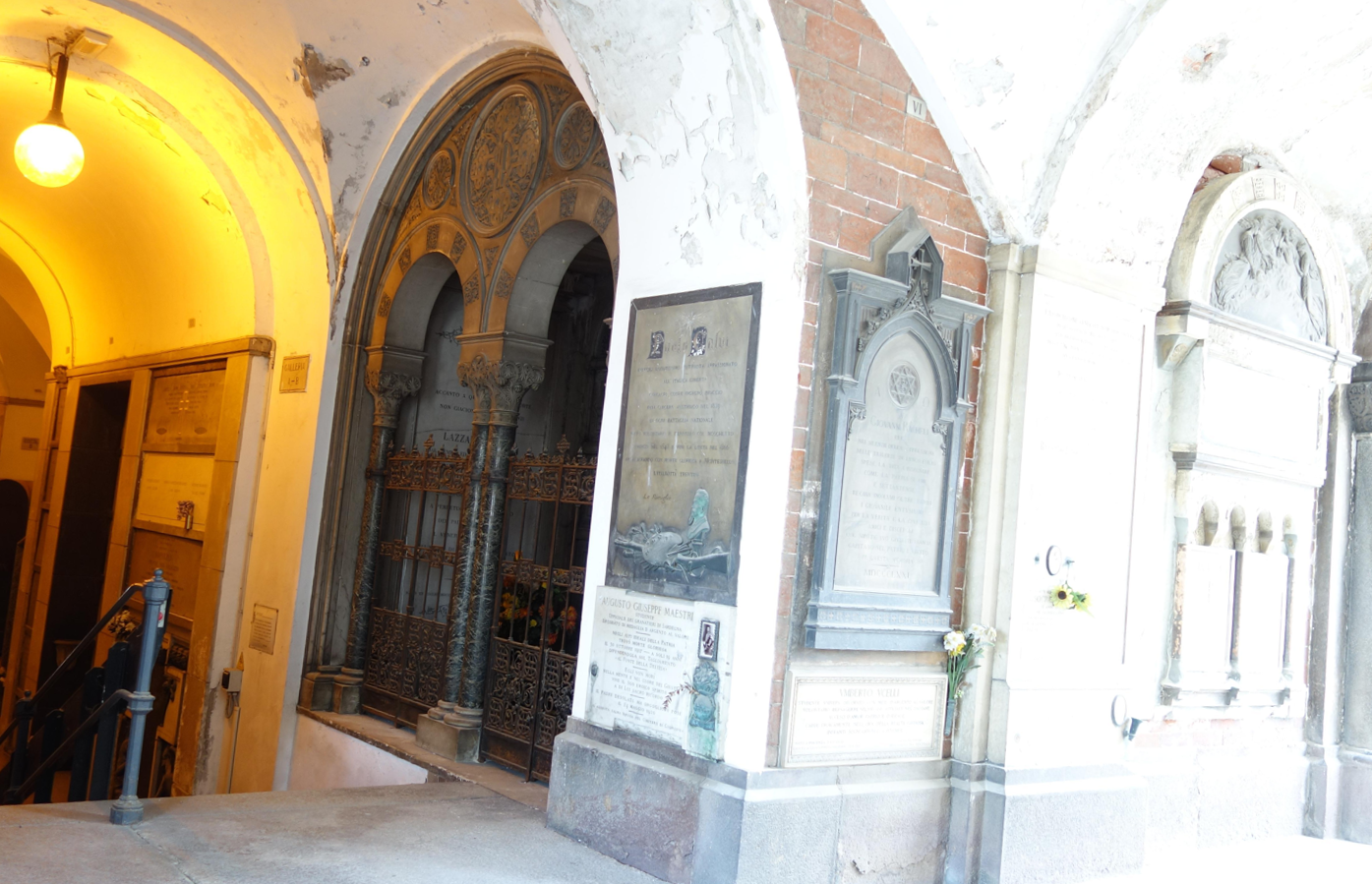}	
		\includegraphics[width=0.108\linewidth]{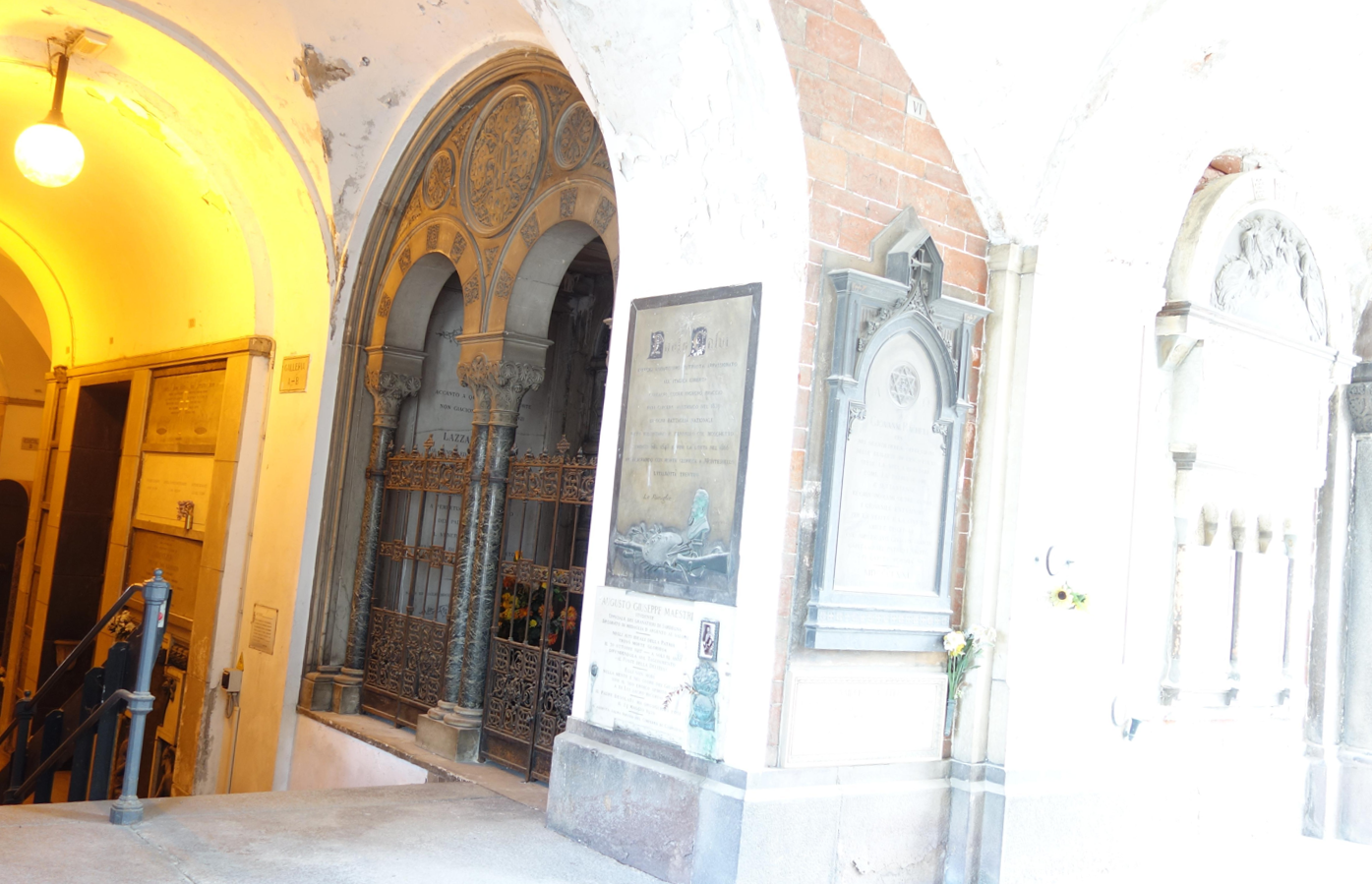}	
		\includegraphics[width=0.108\linewidth]{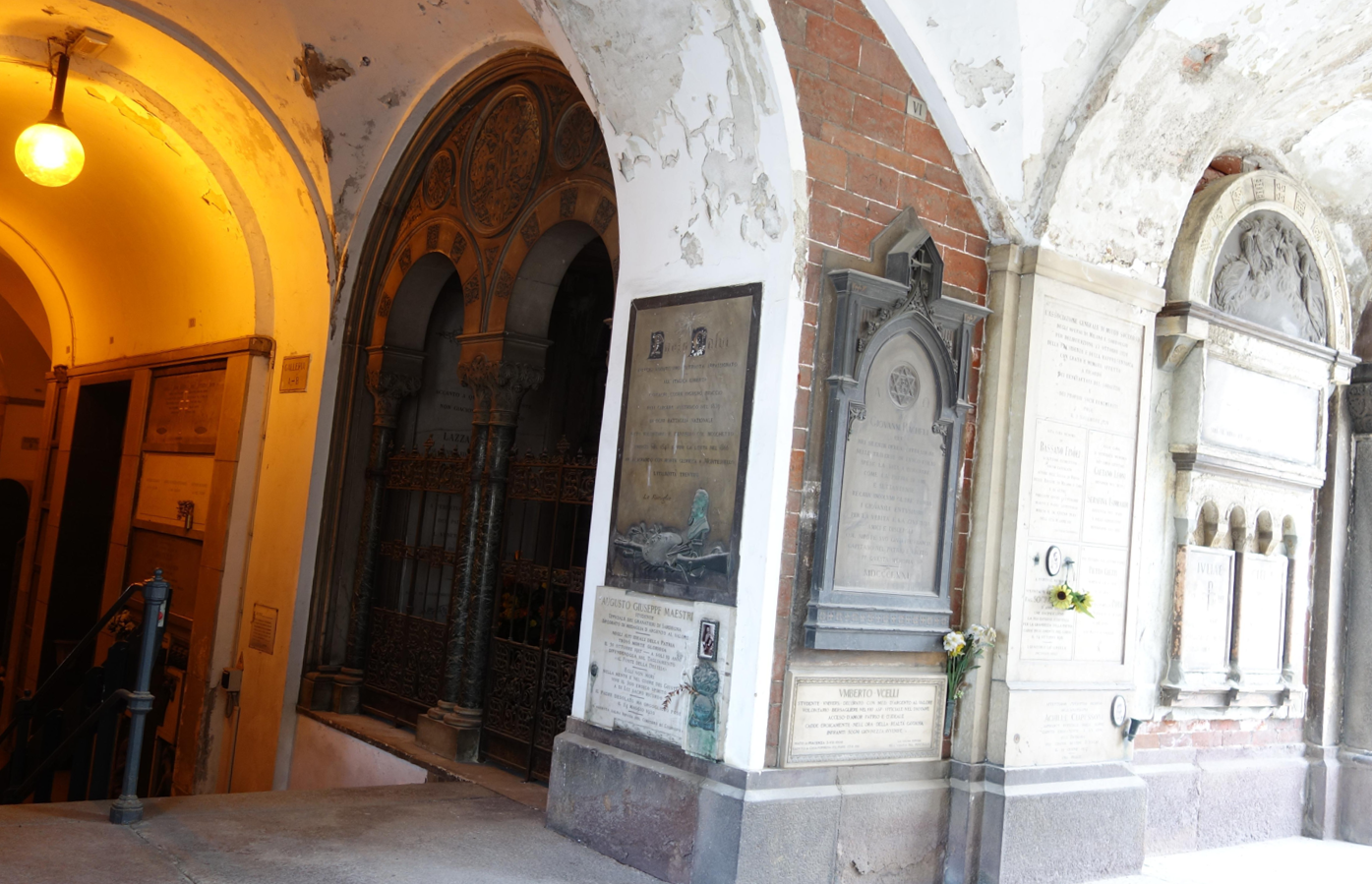}
		\includegraphics[width=0.108\linewidth]{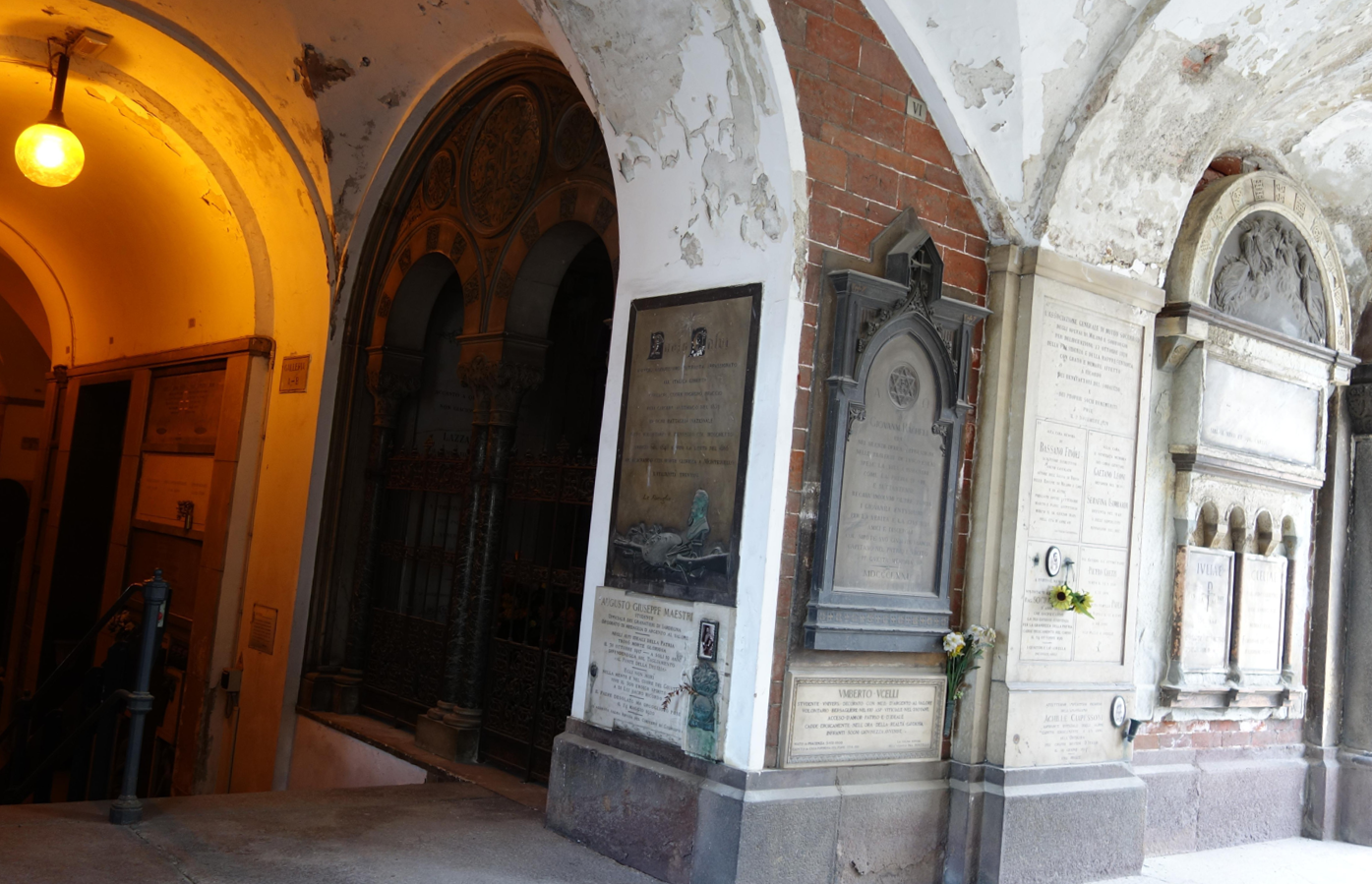}
		\includegraphics[width=0.108\linewidth]{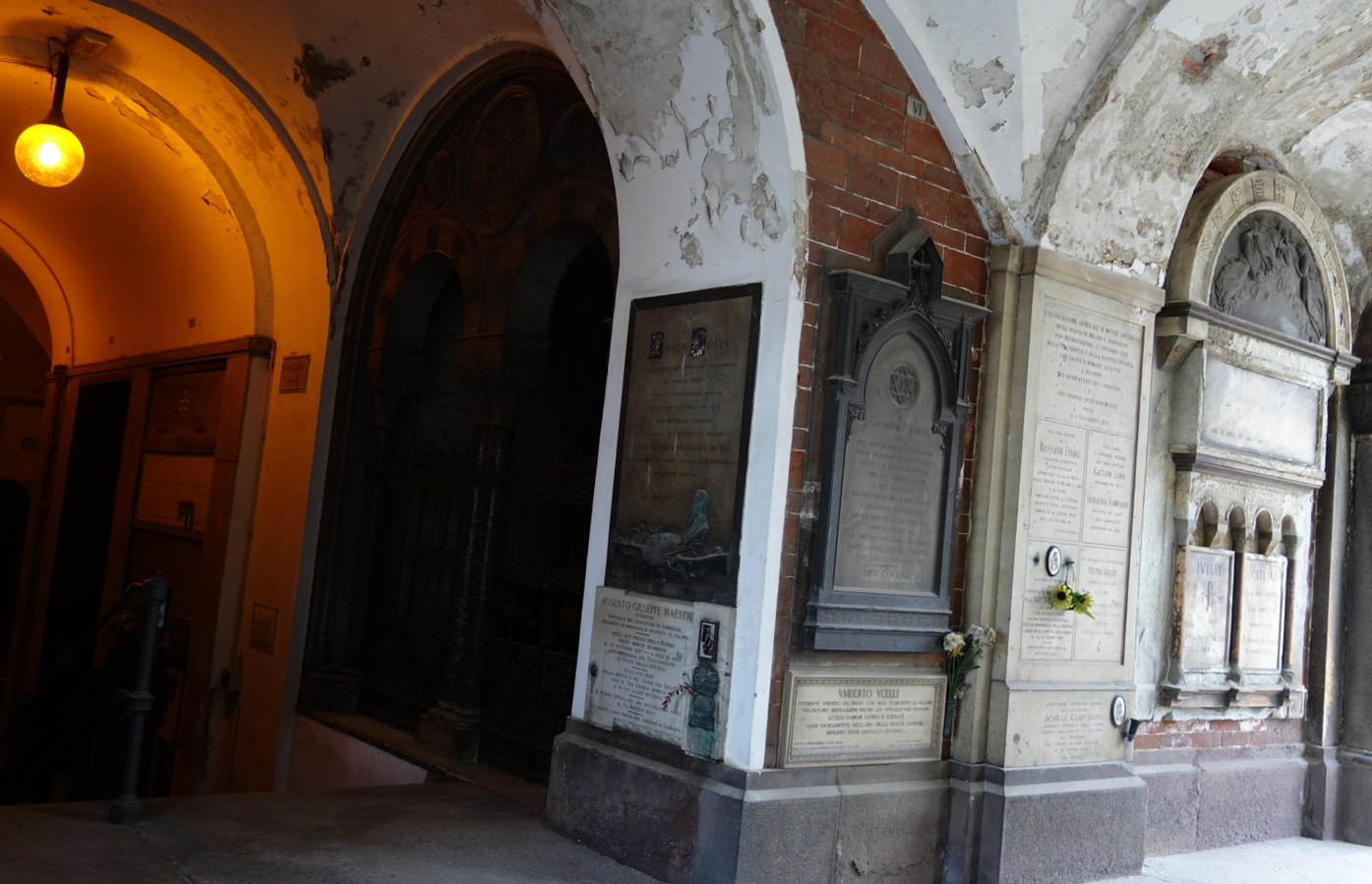}		
		\includegraphics[width=0.108\linewidth]{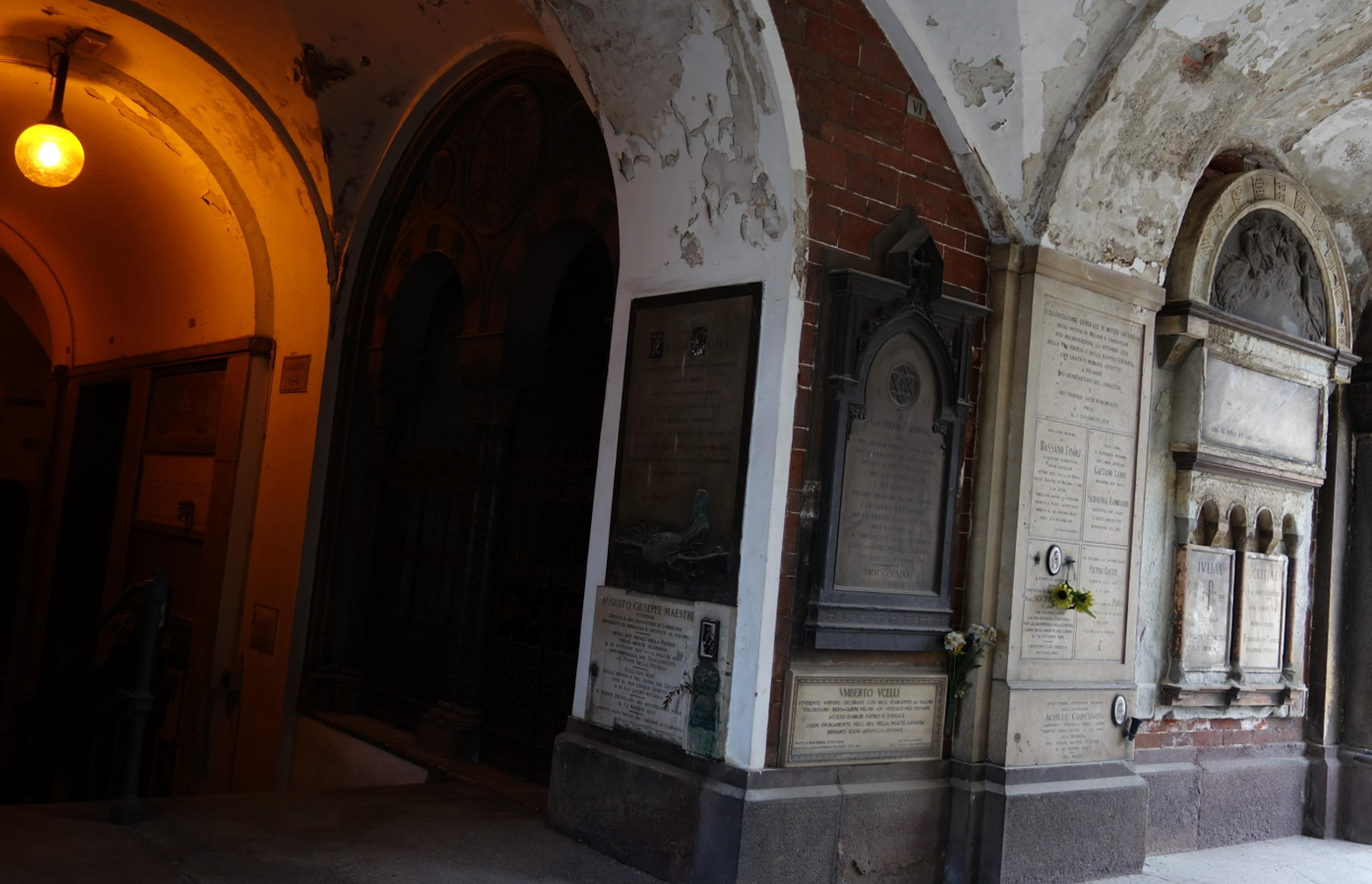} 
		\includegraphics[width=0.108\linewidth]{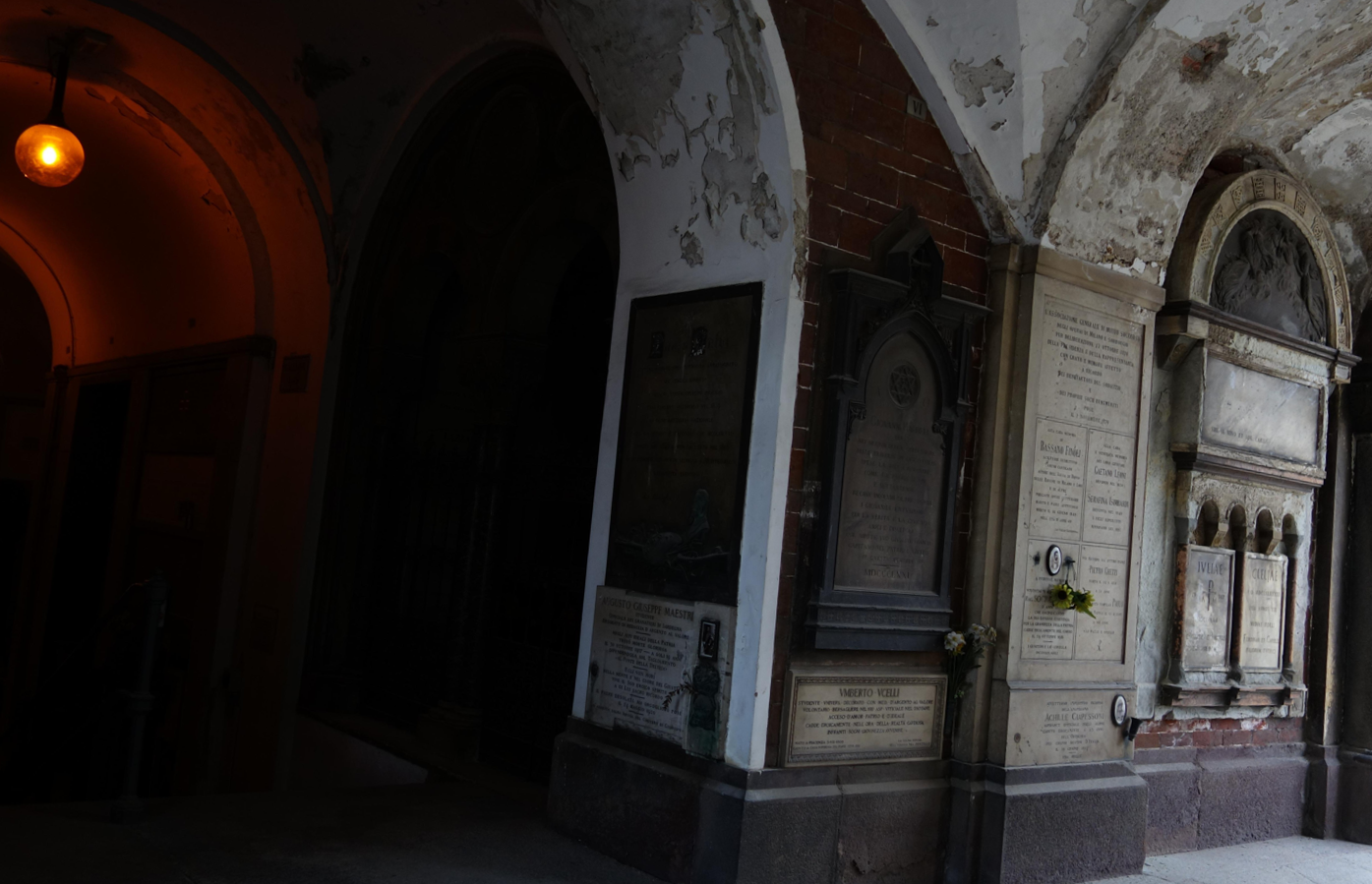} 
		\includegraphics[width=0.108\linewidth]{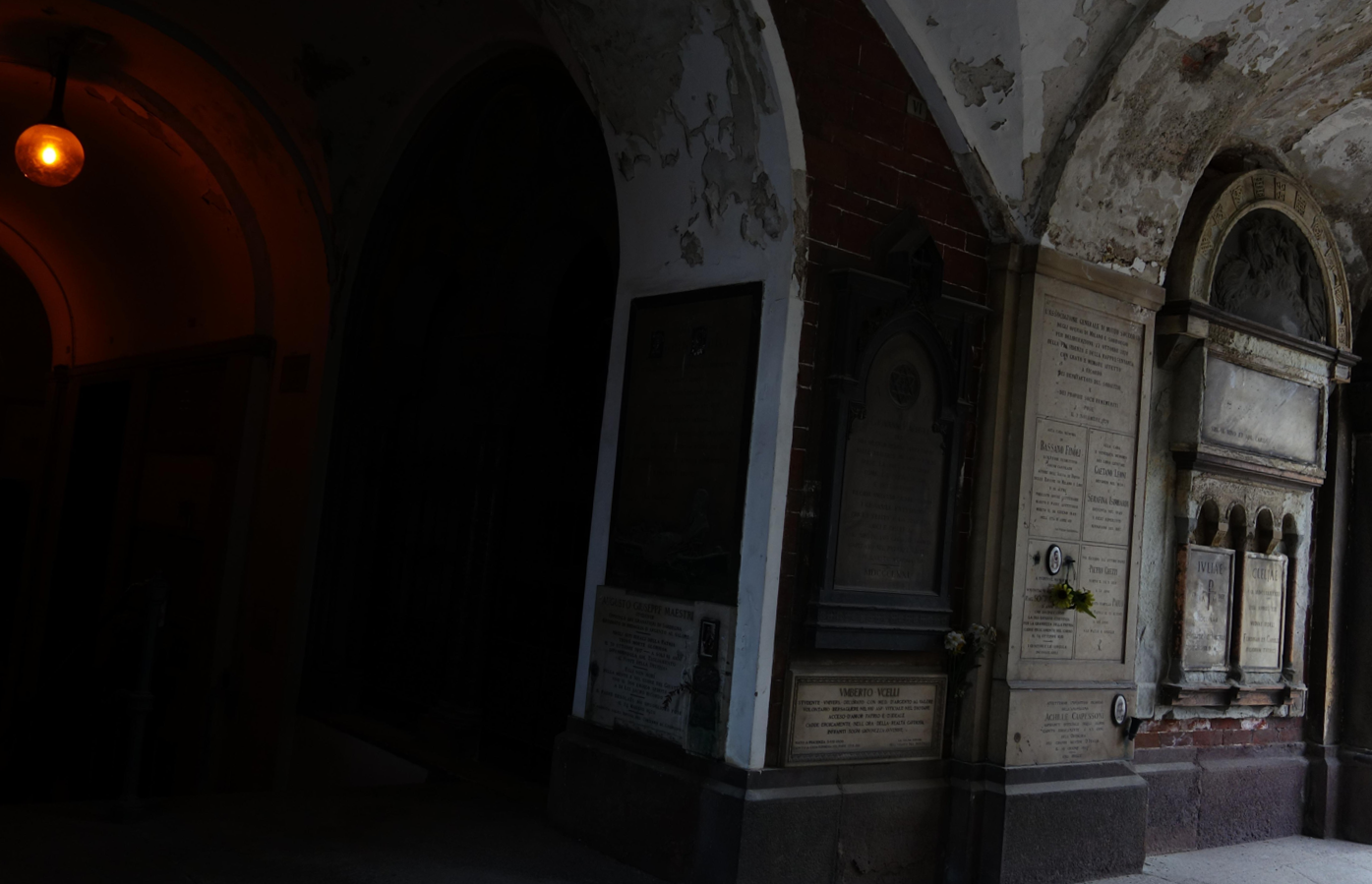}
	}

    \subfigure[``Pillar'' sequence.]{
	   \includegraphics[width=0.122\linewidth]{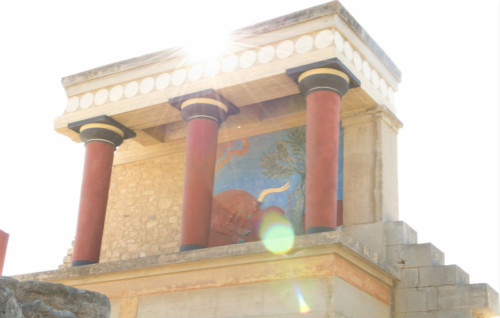} 		
	   \includegraphics[width=0.122\linewidth]{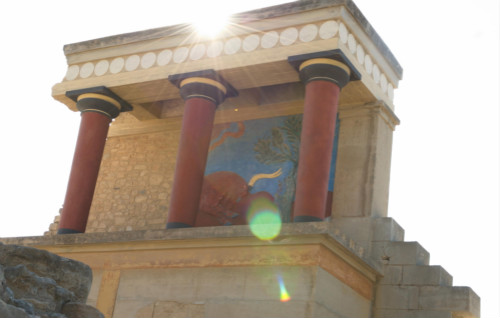}	
	   \includegraphics[width=0.122\linewidth]{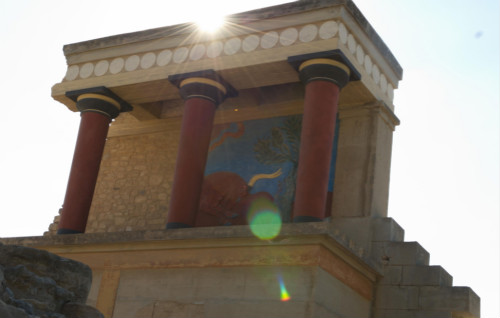}	
	   \includegraphics[width=0.122\linewidth]{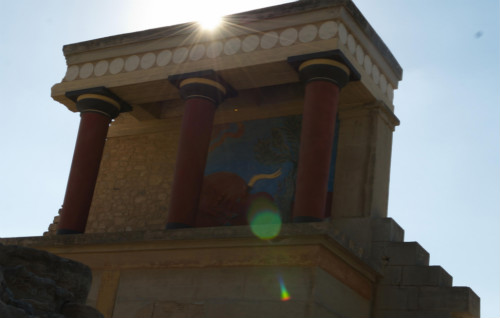}
	   \includegraphics[width=0.122\linewidth]{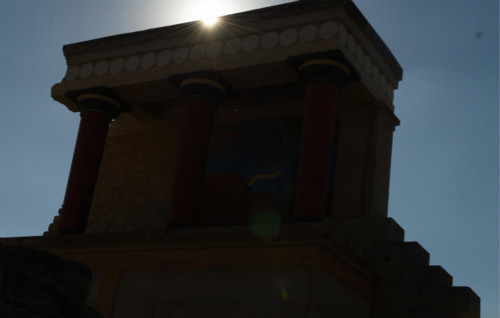}
	   \includegraphics[width=0.122\linewidth]{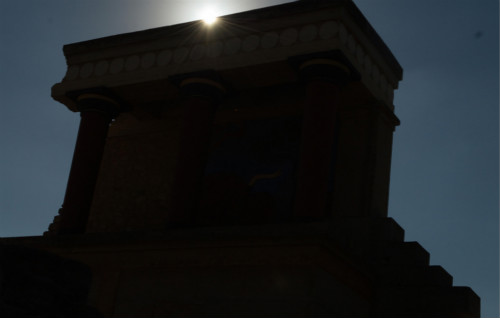}		
	   \includegraphics[width=0.122\linewidth]{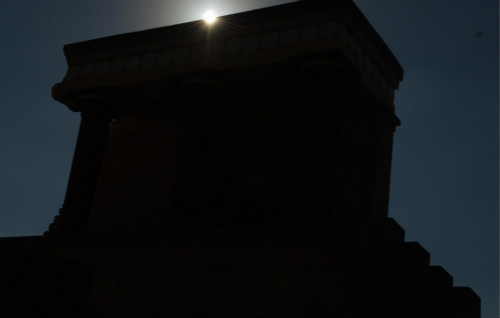} 
	   \includegraphics[width=0.122\linewidth]{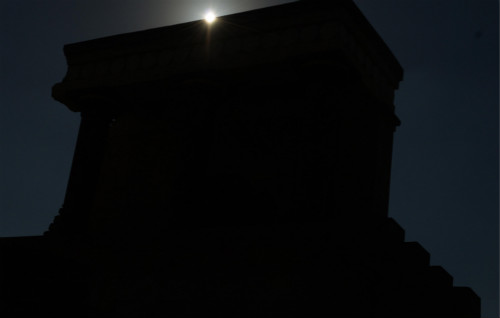}}
	\caption{Multi-exposed image sequences. Each sequence is arranged from long exposure to short exposure and the first image in each sequence is reference image. The interval between images in (a) and (b) is 1EV. The exposure levels in (c) and (d) are manually set based on the lighting ratio of the scene. The full-size images have been uploaded to the internet, readers can zoom them for better observation\cite{data}.}.
	\label{Fig41}
\end{figure*}

9 well geometrically aligned sequences in the benchmark datasets, and 37 randomly selected and well geometrically aligned sequences in \cite{2018DeepSingleImageContrastEnhancer} are used to evaluate different alignment algorithms. These multi-exposure image sequences are captured by different cameras, which cover a broad range of scenes, subjects and lighting conditions. Each sequence has 3 to 18 differently exposed images and is arranged from the longest exposure to the shortest exposure. Some sample sequences are presented in Fig. \ref{Fig41}.

To evaluate the alignment performance, the first image in each sequence is selected as the reference, and other images are rotated by 5$ ^{\circ} $, and shifted 30 pixels and 10 pixels in $y$-axis and $x$-axis, respectively, to produce unaligned samples for latter experiments. The alignment performances are verified in terms of average errors between estimated motion parameters and the true values (5$ ^{\circ} $, 30,10), and the results are shown in Table \ref{tab00} and \ref{tab0}. The average error of the proposed method is the smallest, which implies the proposed method is suitable for various scene sets alignment and robust to intensity variations.

\begin{figure*}
	\centering
	\vspace{-0.35cm}
	{\includegraphics[width=1\textwidth]{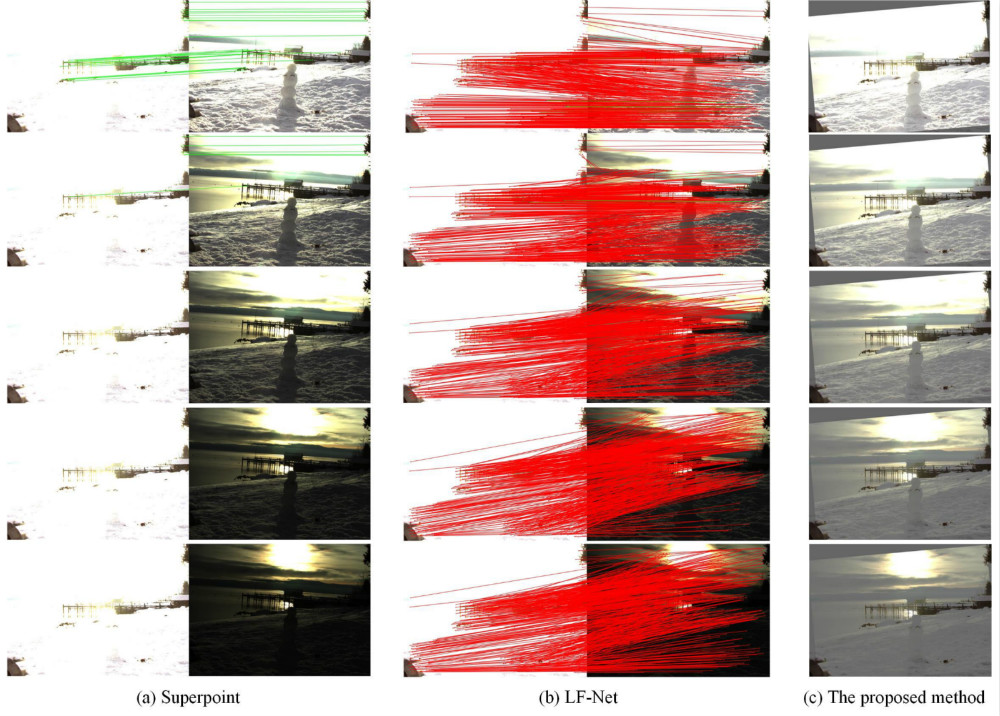} }
	\caption{Comparison of learning-based methods and the proposed method on ``Snowmann'' sequence. In this figure, red lines indicate consistent point pairs, and green lines indicate mismatching point pairs.}
	\label{Fig213}
\end{figure*}

\begin{figure*} 
	\centering
	\vspace{-0.35cm}
	{\includegraphics[width=1\textwidth]{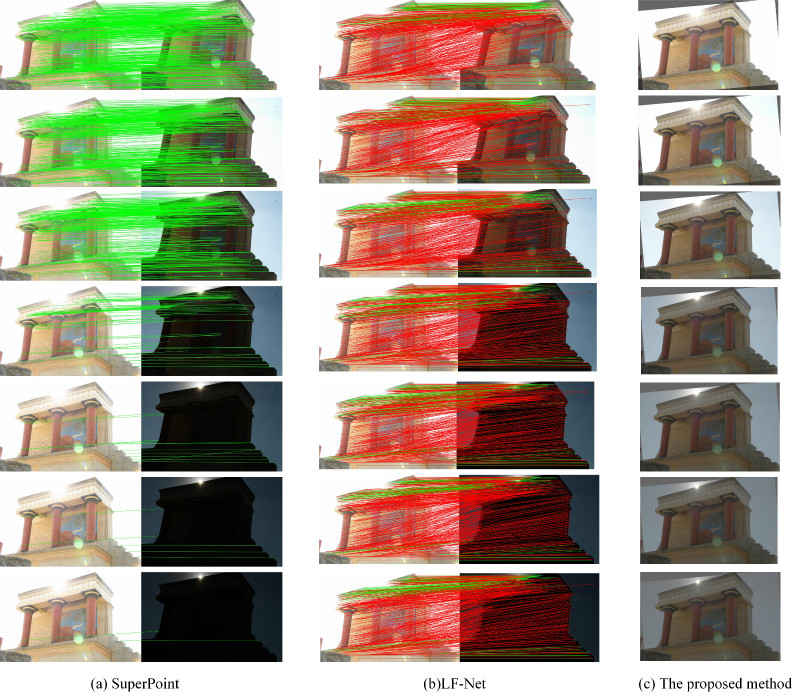} }
	\caption{Comparison of learning-based methods and the proposed method on ``Pillar'' sequence. In this figure, red lines indicate consistent point pairs, and green lines indicate mismatching point pairs.}
   \label{Fig214}
\end{figure*}

A typical sequence named ``Big Tree'' (1EV interval) in benchmark  datasets, shown in Fig. \ref{Fig41}(a), is selected to test each algorithm's robustness to exposure changes. Table \ref{tab1} summarizes the motion errors in aligning for the ``BigTree'' sequence. The SIFT, BRIEF, SuperPoint and  LF-Net do not provide motion parameters but they provide the matching point pairs of two images. Thus, the performances of the four methods are assessed by the number of consistent points and mismatching ones.  Table \ref{tab1} reveals that: 1) Among the order feature-based methods, the error of the MTB method is the smallest, which illustrates that the MTB is robust to the varying exposures. But, as seen in Table \ref{tab2}-\ref{tab21}, the performance of MTB is not good in the other two sequences, such as ``Snowman'' shown in Fig.\ref{Fig41}(b) because the reference image is seriously over-exposed. 2) The performances of the IMF, CT, LBP, SIFT and BRIEF methods are not good, but the IMF+SIFT and IMF+LBP are robust to intensity variations. This shows that IMF improves the similarity of the differently exposed images. 3) BRIEF and IMF+BRIEF do not offer good results due to large intensity variations, while IMF+BRIEF+``HD'' and the proposed method demonstrate high robustness to intensity variations. This indicates that the proposed ``Hamming'' distance improves the alignment accuracy of binary operators, especially for the rotational angle $\theta$. The proposed normalization method reduces the effects of saturation and noise in the dark regions, and improves the similarity of two differently exposed images.  4) It can be observed from Table \ref{tab1} and Table \ref{tab2} that the performance of the learning based methods on ``Snowman'' sequence is worse than their performance on ``BigTree'' sequence because the reference image on ``Snowman'' sequence is completely washed out. The SuperPoint is more robust to intensity variations than the LF-Net, i.e., it can find 6 pairs of consistent points in the third image, but 0 pair from the fourth image due to large exposure difference between them, as shown in Fig. \ref{Fig213}(a). By contrast, the method of LF-net can find more feature points, but can not match these feature points correctly, as shown in Fig. \ref{Fig213}(b).  5) In comparison with the above methods, the proposed method is robust to large EV intervals (7EV in ``Bigtree''), especially for over-exposed image sequences, in which the reference image is completely washed out. The LBP based alignment algorithm does not work as indicated in Table \ref{tab2}. This was also pointed out in \cite{1alis2017}. The performance of SuperPoint on sequences ``Pillar'' and  ``Inscriptionn'' is better than its performance on ``Snowman'' and ``BigTree'' sequences because the reference images in sequences ``Pillar'' and  ``Inscriptionn'' contain rich features as shown in Fig. \ref{Fig214}. Nevertheless, the proposed method is also superior to other methods as shown in Table \ref{tab21} and Table \ref{tab22}. The proposed image alignment algorithm outperforms the DVO algorithm in \cite{1alis2017} and multi-exposed image alignment in \cite{1wu2014}.

\begin{table}
	\centering	
	\caption{Overall results performed on benchmark database. $\Delta \theta$, $\Delta t_{y}$ and $\Delta t_{x}$, are the average errors between estimated motion parameters and (5$ ^{\circ} $, 30, 10). }
	\label{tab00}
	\setlength{\tabcolsep}{3pt}
	\footnotesize  
	\begin{tabular}{|p{40pt}|c|c|c|c|c|c|c|c|c|}
		\hline 
		\multirow{2}*{Method} &\multicolumn{3}{c|} {$\Delta\theta$($ ^{\circ} $)}  &\multicolumn{3}{c|}{$\Delta t_{y}$}     &\multicolumn{3}{c|}{$\Delta t_{x}$}\\    
		\cline{2-10}&Mean&Max&Min  		&Mean&Max&Min    	&Mean&Max&Min\\		
		\hline 
		MTB &2.8&23.5&0  &17.8&149&0.8   &9.6&92&0.4 \\ 
		CT  &6&39&0  &64&539&\textbf{0}   &49&195&0.5 \\ 
		LBP &3&23&0  &26&213&0.8  &34&282&0 \\ 
		IMF &1.3&\textbf{6.2}&0  &4.1&\textbf{32}&0.1  &6.1&130&0 \\ 
		IMF+LBP &1.1&13&0  &8.5&192&0.9  &10&254&0 \\ 
		
		IMF+BRIEF
		+``HD''   &\multirow{2}*{1.9} &\multirow{2}*{14.1}&\multirow{2}*{0}  &\multirow{2}*{14.6} &\multirow{2}*{181} &\multirow{2}*{0.3}   &\multirow{2}*{15.8}&\multirow{2}*{253}&\multirow{2}*{0.3}\\ 
		Ours &\textbf{0.6}&10&0  &\textbf{3.8}&34&0.7  &\textbf{1.8}&\textbf{11.1}&0.1  \\
		\hline
	\end{tabular}
\end{table}

\begin{table}
	\centering		
	\caption{Overall results performed on 37 Cai's database\cite{2018DeepSingleImageContrastEnhancer}.}
	\label{tab0}
	\setlength{\tabcolsep}{3pt}
	\footnotesize
	\begin{tabular}{|p{40pt}|c|c|c|c|c|c|c|c|c|}
		\hline 
		\multirow{2}*{Method} &\multicolumn{3}{c|} {$\Delta\theta$($ ^{\circ} $)}  &\multicolumn{3}{c|}{$\Delta t_{y}$}     &\multicolumn{3}{c|}{$\Delta t_{x}$}\\    
		\cline{2-10}&Mean&Max&Min  		&Mean&Max&Min    	&Mean&Max&Min\\		
		\hline 
		MTB      &0.9&23&0     &7.2&177&0.1  &4.2&151&0\\ 
		
		CT       &5.2&62&0     &86&542&0.5   &155&984&0 \\ 
		
		LBP      &0.7&12&0      &11&\textbf{137}&0.1   &8.3&369&0  \\ 
		
		IMF      &1.6&6&0      &7.1&201&0.1  &7.3&386&0\\ 
		
		IMF+LBP  &0.9&24&0     &8.8&195&0    &8.3&346&0\\ 
		
		IMF+BRIEF
		+``HD''  &\multirow{2}*{0.7}&\multirow{2}*{14.8}&\multirow{2}*{0}         &\multirow{2}*{11.2} &\multirow{2}*{199} &\multirow{2}*{0}     &\multirow{2}*{9.1} &\multirow{2}*{352} &\multirow{2}*{0}\\ 
		
		Ours     &\textbf{0.2}&\textbf{3.6}&0   &\textbf{6.9}&241&0  &\textbf{1.5}&\textbf{9.9}&0\\ 
		\hline		
	\end{tabular} 
\end{table}

\begin{table*}
	\centering
	\vspace{-0.35cm}
	\caption{Performances of different methods on ``BigTree'' sequence in terms of motion parameters $(\Delta\theta, \Delta t_{y},\Delta t_{x})$.}
    \setlength{\tabcolsep}{3pt}
    \footnotesize    
	\begin{tabular}{|c|c|c|c|c|c|c|c|}
		\hline
		\multicolumn{2}{|c|}{Method}                & 2  & 3  & 4  & 5  & 6  & 7\\
		\hline
		\multirow{2}*{MTB}      &$\Delta \theta$ &0 &0  & 0  &0 &0  &\textbf{0}\\  
		&$\Delta t_{y},\Delta t_{x}$         & 1.6,1.1	&1.6,0.9	&1.7,0.8	&1.7,0.7	&1.7,0.7	&\textbf{1.6,0.7}\\			
		\multirow{2}*{CT}       &$\Delta \theta$    & 0  & 0  & 1.8	& 1.4 & 3  & 5\\
		&$\Delta t_{y},\Delta t_{x}$         & 1,1.1	    &0.6,1.2	&34,19	    &84,175	    &21,195	    &59,192\\			
		\multirow{2}*{LBP}       &$\Delta \theta$   &0	&0	&0.5	&1	&3.2	&11.7\\
		&$\Delta t_{y},\Delta t_{x}$         &2,1	&2,1	&5.5,0.6	&19,138	&17,33	&36,65\\			
		\multirow{2}*{IMF}      &$\Delta \theta$    &2.4	&1.6	&1.9	&3	&2.6	&\textbf{0}\\
		&$\Delta t_{y},\Delta t_{x}$         &0.4,3	&0.1,2.2	&0.2,2.3	&2,5.5	&1.1,5.1	&\textbf{1.8,1.2}\\			
		\multirow{2}*{IMF+LBP}  &$\Delta \theta$    & 0  &0  &0  &0  &0	&4.4\\
		&$\Delta t_{y},\Delta t_{x}$         &1.7,1.3	&1.7,1.3	&1.7,1.3	&1.6,1.3	&1.7,1.3	&15,11\\			
		\multirow{2}*{IMF+BRIEF+``HD''}   &$\Delta \theta$      &0  &0  & 0  &0  &0&6\\
		&$\Delta t_{y},\Delta t_{x}$         &3,1.2	 &2,5.5	  &2.3,0	&1,5   &2.7,4.3	&31,12\\			
		\multirow{2}*{Ours}       &$\Delta \theta$  &0  &0  & 0  &0  &0 &\textbf{0}\\
		&$\Delta t_{y},\Delta t_{x}$         &1.6,1.4	&1.6,1.4	&1.6,1.4	&1.6,1.5	&3.2,0	&\textbf{1.5,1.5}\\			
		\hline
		\multicolumn{2}{|c|}{SIFT}      &$\surd $ 	&$\surd $	&$\times$	&$\times$	&$\times$	&$\times$ \\	
     	\multicolumn{2}{|c|}{IMF+SIFT}  &$\surd $ 	&$\surd$	&$\surd$	&$\surd $	&$\times$	&$\times$ \\	
     	\multicolumn{2}{|c|}{BRIEF}     &$\times$ 	&$\times$	&$\times$	&$\times$	&$\times$	&$\times$ \\	
     	\multicolumn{2}{|c|}{IMF+BRIEF} &$\times $ 	&$\times$	&$\times$	&$\times$	&$\times$	&$\times$ \\
     	\multicolumn{2}{|c|}{SuperPoint}  &$\surd $ 	&$\surd$	&$\surd$	&$\times$	&$\times$	&$\times$ \\     
     	\multicolumn{2}{|c|}{LF-Net}  &$\times $ 	&$\times$	&$\times$	&$\times$	&$\times$	&$\times$ \\	
     \hline 
     \multicolumn{8}{c}{$\surd $ indicates good alignment, $\times $indicates wrong alignment.}         
	\end{tabular}
	\label{tab1}
\end{table*}

\begin{table*}
	\centering
	\caption{Performances of different methods on ``Snowman'' sequence in terms of motion parameters $(\Delta\theta, \Delta t_{y},\Delta t_{x})$.}
    \setlength{\tabcolsep}{3pt}
    \footnotesize    
	\begin{tabular}{|c|c|c|c|c|c|c|}
		\hline
		\multicolumn{2}{|c|}{Method}                                   & 2  & 3  & 4  & 5  & 6  \\
		\hline
		\multirow{2}*{MTB}       &$\Delta \theta$                     & 0.7  & 8.5  & 10  & 10  & 8.6 \\  
		&$\Delta t_{y},\Delta t_{x}$         & 3,1	&32,4	&38,15	&40,17	&33,4\\			
		\multirow{2}*{CT}        &$\Delta \theta$                     & 0.2  & 1.6  & 7	& 20 & 39 \\
		&$\Delta t_{y},\Delta t_{x}$         & 0,1	    &32,150 	&2,77	    &39,38	    &20,75\\			
		\multirow{2}*{LBP}       &$\Delta \theta$                     &3	&5	&7.4	&5.4	&0.4\\
		&$\Delta t_{y},\Delta t_{x}$         &17,34	&51,130	    &0,189	    &31,145	    &157,127\\		
		\multirow{2}*{IMF}       &$\Delta \theta$                     &6	&6	&5.4	&2.4	&5.2\\
		&$\Delta t_{y},\Delta t_{x}$         &16,3	&16,3	 &30,9	&32,10	 &17,10\\			
		\multirow{2}*{IMF+LBP}   &$\Delta \theta$                     &0	&0	&0	&0.1	&\textbf{0.1}\\
		&$\Delta t_{y},\Delta t_{x}$         &1.4,0.8	&1.6,0.7	&1.4,1	&1.4,1	&\textbf{1.5,1}\\		
		\multirow{2}*{IMF+BRIEF+``HD''}   &$\Delta \theta$      &1.3	&0.5	&0.1	&0.7	&0.7\\
		&$\Delta t_{y},\Delta t_{x}$         &11,5	 &1,0	  &15,16	&5,6   &11,10\\
		\multirow{2}*{Ours}       &$\Delta \theta$   &0 &0  &0 &0  &\textbf{0}\\
		&$\Delta t_{y},\Delta t_{x}$         &1.6,1.1	&1.7,1.0  &1.8,1.0	&1.8,1.0	&\textbf{1.8,0.8}\\			
		\hline
		\multicolumn{2}{|c|}{SIFT}  &$\surd $ 	&$\surd $	&$\times$	&$\times$	&$\times$ \\	
	
		\multicolumn{2}{|c|}{IMF+SIFT}  &$\surd $ 	&$\surd$	&$\surd$	&$\surd $	&$\surd$\\	
		
		\multicolumn{2}{|c|}{BRIEF} &$\times$ 	&$\times$	&$\times$	&$\times$	&$\times$\\	
		
		\multicolumn{2}{|c|}{IMF+BRIEF}  &$\times $ 	&$\times$	&$\times$	&$\times$	&$\times$\\	
		\multicolumn{2}{|c|}{SuperPoint}  &$\surd $ 	&$\surd$	&$\times$	&$\times$	&$\times$ \\	
		\multicolumn{2}{|c|}{LF-Net}  &$\times $ 	&$\times$	&$\times$	&$\times$	&$\times$ \\
		\hline 
		\multicolumn{7}{c}{$\surd $ indicates good alignment, $\times $indicates wrong alignment.}                
	\end{tabular}
	\label{tab2}
\end{table*}

\begin{table*}
	\centering
	\vspace{-0.35cm}
	\caption{Performances of different methods on ``Pillar'' sequence in terms of motion parameters $(\Delta\theta, \Delta t_{y},\Delta t_{x})$.}
	\label{tab21}
	\setlength{\tabcolsep}{3pt}
	\footnotesize    
	\begin{tabular}{|c|c|c|c|c|c|c|c|c|}
		\hline
		\multicolumn{2}{|c|}{Method}              & 2  & 3  & 4  & 5  & 6 &7 &8 \\
		\hline
		\multirow{2}*{MTB}    &$\Delta \theta$    &0.1  &0.1 &0.1 &0.1 &0.1  &0.3  &20.5 \\  
		&$\Delta t_{y},\Delta t_{x}$         &2.1,1.2 &1.6,1.5 &1.3,1.9 &1.4,1.5 &0.8,1.4 &0.1,0.7 &156.5,1.7\\			
		\multirow{2}*{CT}    &$\Delta \theta$     &0 & 0 &0 &4.7 &3.6 &3.3 &6 \\
		&$\Delta t_{y},\Delta t_{x}$         &2,0.8&1.6,1.1 &1.2,1.5 &89,20 &470,86 &471,115 &205,595\\			
		\multirow{2}*{LBP}       &$\Delta \theta$  &0 &0  &0 &0.1 &0.1 &6.3 &5.7\\
		&$\Delta t_{y},\Delta t_{x}$         &2,0.8 &1.6,1.1 &1.2,1.5&1.4,1.1&1.6,0.8&47,16.2&46.3,29\\		
		\multirow{2}*{IMF}  &$\Delta \theta$    &4.3 &4.6 &2.2 &0.8 &0.6 &0.4 &\textbf{0.5}\\
		&$\Delta t_{y},\Delta t_{x}$   &9.7,7.7 &12,9 &11,6 &3.4,2.7 &2.5,2.2 &1.2,0.8 &\textbf{1.3,0.2}\\			
		\multirow{2}*{IMF+LBP}   &$\Delta \theta$   &0	&0	&0 &0	&0 &0.2 &3\\
		&$\Delta t_{y},\Delta t_{x}$     &2,0.9 &1.6,1.2 &1.3,1.6 &1.4,1.5 &1.5,1.4 &1.7,0.7 &15,6.1\\		
		\multirow{2}*{IMF+BRIEF+``HD''}   &$\Delta \theta$ &0.2 &0.1 &0.6 &0 &1.2 &0.4 &0.5\\
		&$\Delta t_{y},\Delta t_{x}$      &3.9,12.6 &0,7.1 &9.8,16.3 &5.1,5.5 &1.6,30 &33.6,6.7 &12.7,8.9\\
		\multirow{2}*{Ours}       &$\Delta \theta$   &0  &0 &0 &0 &0&0.1&\textbf{0.1}\\
		&$\Delta t_{y},\Delta t_{x}$    &2,0.9 &1.6,1.2 &1.2,1.6 &1.4,1.4 &1.5,1.2 &1.2,0.4 &\textbf{1.2,0.2}\\			
		\hline
		\multicolumn{2}{|c|}{SIFT}  &$\surd $ 	&$\times $	&$\times$	&$\times$	&$\times$ &$\times $ &$\times $\\	
		
		\multicolumn{2}{|c|}{IMF+SIFT}  &$\surd $ 	&$\times $	&$\times$	&$\times$	&$\times$ &$\times $ &$\times $\\	
		
		\multicolumn{2}{|c|}{BRIEF} &$\times$ 	&$\times$	&$\times$	&$\times$	&$\times$ &$\times $ &$\times $\\	
		
		\multicolumn{2}{|c|}{IMF+BRIEF}  &$\surd $ 	&$\times$	&$\times$	&$\times$	&$\times$&$\times $ &$\times $\\	
		\multicolumn{2}{|c|}{SuperPoint}  &$\surd $ 	&$\surd $	&$\surd $	&$\surd $	&$\surd $ &$\surd $ &$\times $\\	
		\multicolumn{2}{|c|}{LF-Net}  &$\times $ 	&$\times$	&$\times$	&$\times$	&$\times$ &$\times$ &$\times$ \\
		\hline 
		\multicolumn{7}{c}{ $\surd $ indicates good alignment, $\times $indicates wrong alignment.}                
	\end{tabular}
\end{table*}

\begin{table*}
	\centering
	\caption{Performances of different methods on ``Inscription'' sequence in terms of motion parameters $(\Delta\theta, \Delta t_{y},\Delta t_{x})$.}
	\setlength{\tabcolsep}{3pt}
	\footnotesize    
	\begin{tabular}{|c|c|c|c|c|c|c|c|c|c|}
		\hline
		\multicolumn{2}{|c|}{Method}              & 2  & 3  & 4  & 5  & 6 &7 &8 &9 \\
		\hline
		\multirow{2}*{MTB}    &$\Delta \theta$    &0.3  &0.1 &0 &0 &0  &0.1  &0.1  &\textbf{0.1} \\  
		&$\Delta t_{y},\Delta t_{x}$         &3.8,1.2 &1.0,1.4 &0.8,1.5 &0.4,1.8&0.5,1.7&1.1,1.4&1.2,1.5&\textbf{1.8,1.3}\\			
		\multirow{2}*{CT}    &$\Delta \theta$     &0 & 0.1 &6 &6.8  &7.4  &3.6	&2.6 &7 \\
		&$\Delta t_{y},\Delta t_{x}$         &0.9,1 &0.8,1.1 &15,233& 48,4 &82.5,523&51,570&36,675&63,707\\			
		\multirow{2}*{LBP}       &$\Delta \theta$  &0 &0   &0.1 &0.1 &0.1 &0 &0 &\textbf{0}\\
		&$\Delta t_{y},\Delta t_{x}$         &1.3,1.2&1.1,1.3&1.0,1.1&0.6,1.4&0.8,1.2&1.4,0.8&1.4,0.7&\textbf{1.8,0.5}\\		
		\multirow{2}*{IMF}       &$\Delta \theta$    &2.2 &4 &2.1 &3.2 &3.3 &3.3 &3.5 & 0.1\\
		&$\Delta t_{y},\Delta t_{x}$         &4.8,6.6 &23,7.5 &7.4,6.6 &11.4,7.9 &11.5,8 &11,7.9 &8.4,9.5 &2.2,0.9\\			
		\multirow{2}*{IMF+LBP}   &$\Delta \theta$ &0	&0	&0.1 &0.1	&0.1 &0 &0	&24\\
		&$\Delta t_{y},\Delta t_{x}$     &1.3,1.4 &1.1,1.4 &1,1.5 &0.7,1.8 &0.8,1.7&1.4,1.3 &1.4,1.3 &190,333\\		
		\multirow{2}*{IMF+BRIEF+``HD''}   &$\Delta \theta$  &0 &0 &0 &0.2 &0.1 &0.1 &0 &14.8\\
		&$\Delta t_{y},\Delta t_{x}$      &4.3,17 &1.2,2 &4.6,20.5 &1.5,0.3 &0.9,2.3 &4.5,14.4 &0.6,13.5 &27,11\\
		\multirow{2}*{Ours}       &$\Delta \theta$  &0  &0 &0.1 &0.1 &0.1	&0 &0  &\textbf{0} \\
		&$\Delta t_{y},\Delta t_{x}$    &1.2,1.3 &1.1,1.4 &1,1.4 &0.5,1.6 &0.7,1.6 &1.3,1.1 &1.3,1.2&\textbf{1.7,1.1}\\			
		\hline
		\multicolumn{2}{|c|}{SIFT}  &$\surd $ 	&$\surd $	&$\surd $ &$\surd $ &$\surd $ &$\surd $	&$\times$	&$\times$ \\	
		
		\multicolumn{2}{|c|}{IMF+SIFT}  &$\surd $ 	&$\surd $	&$\surd $ &$\surd $ &$\surd $ &$\surd $	&$\surd $	&$\times$\\	
		
		\multicolumn{2}{|c|}{BRIEF} &$\times$ 	&$\times$	&$\times$	&$\times$	&$\times$ &$\times$ 	&$\times$	&$\times$\\	
		
		\multicolumn{2}{|c|}{IMF+BRIEF}  &$\times $ 	&$\times$	&$\times$	&$\times$	&$\times$ &$\times$ 	&$\times$	&$\times$\\	
		\multicolumn{2}{|c|}{SuperPoint}  &$\surd $ 	&$\surd $	&$\surd $ &$\surd $ &$\surd $ &$\surd $	&$\surd $	&$\surd $\\	
		\multicolumn{2}{|c|}{LF-Net}  &$\times $ 	&$\times$	&$\times$	&$\times$	&$\times$ &$\times$	&$\times$	&$\times$ \\
		\hline 
		\multicolumn{7}{c}{ $\surd $ indicates good alignment, $\times $indicates wrong alignment.}                
	\end{tabular}
	\label{tab22}
\end{table*}

\subsection{Test on real images}

\begin{figure*}
	\centering	
	\vspace{-0.35cm}	
	\subfigure[``Snow'' sequence. ]{
		\includegraphics[width=0.25\linewidth]{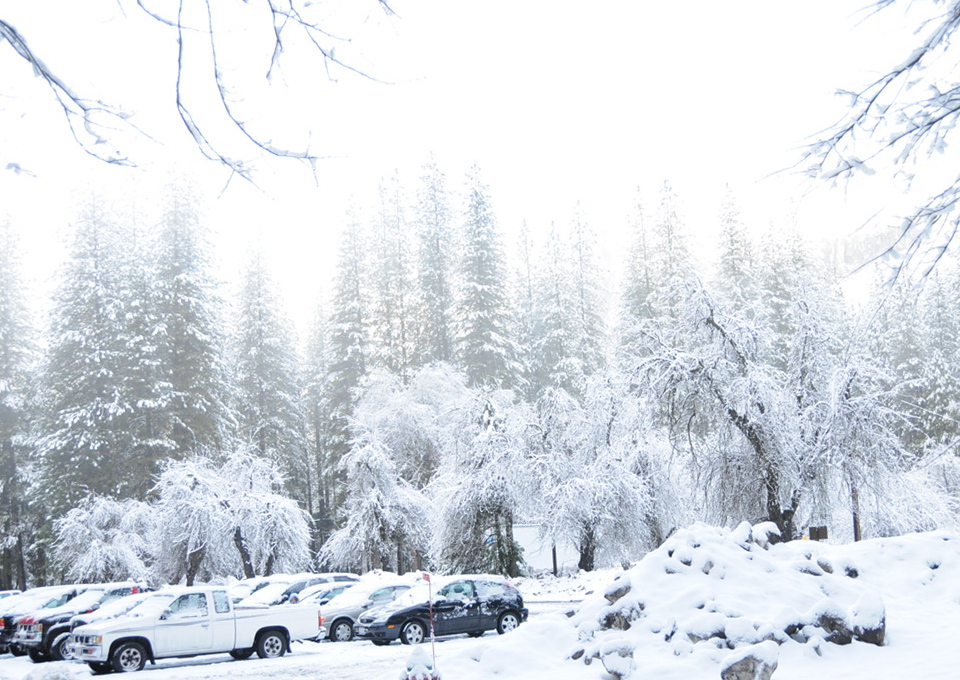}
		\includegraphics[width=0.25\linewidth]{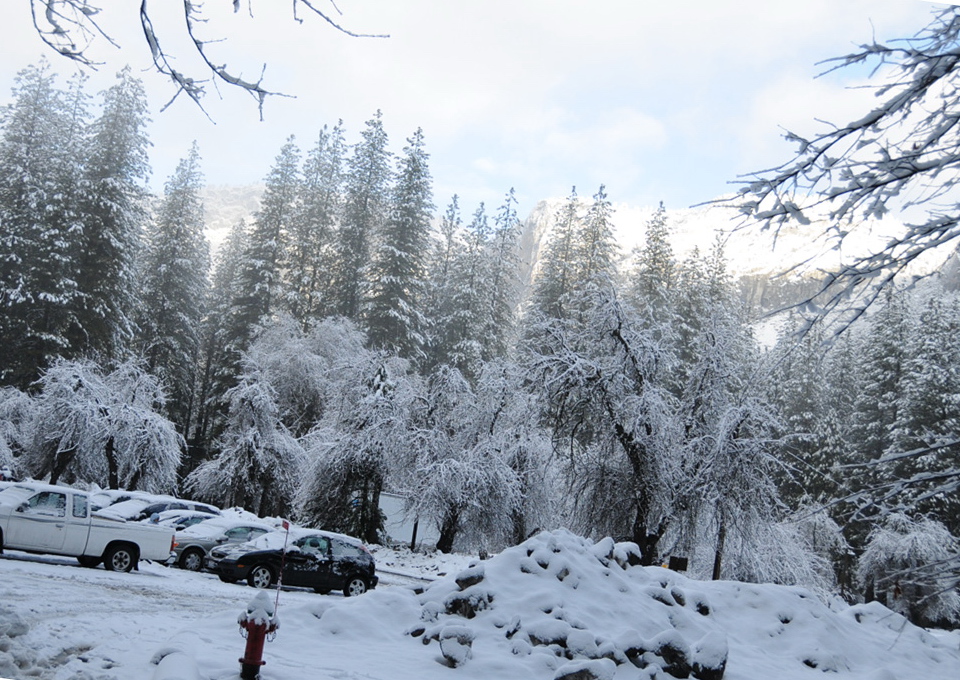}
		\includegraphics[width=0.25\linewidth]{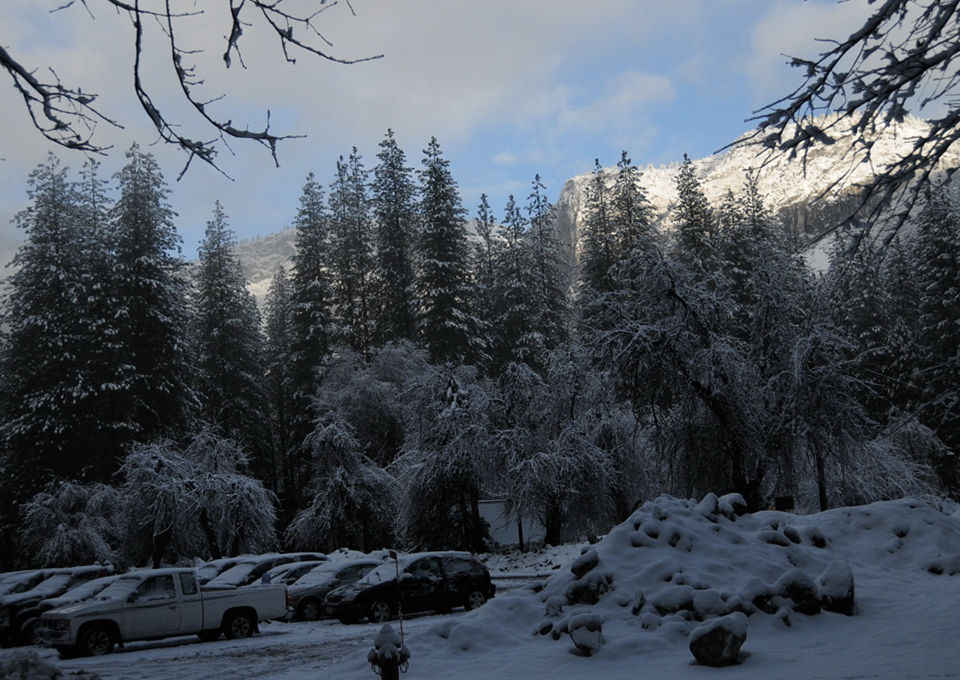}
		\includegraphics[width=0.25\linewidth]{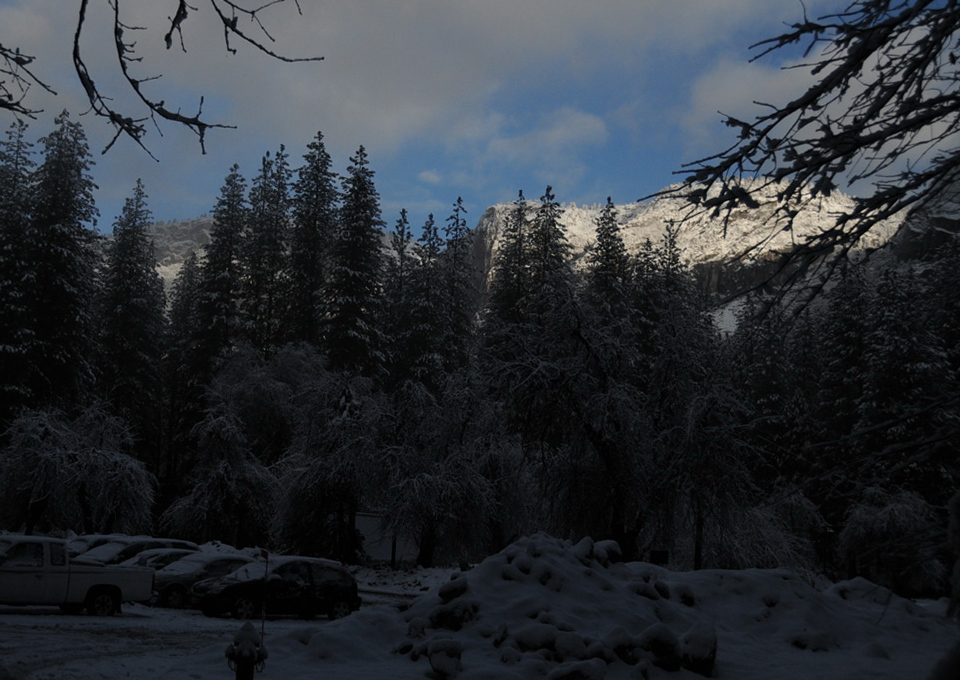}}
	
	\subfigure[``Table'' sequence.]{		
		\includegraphics[width=0.108\linewidth]{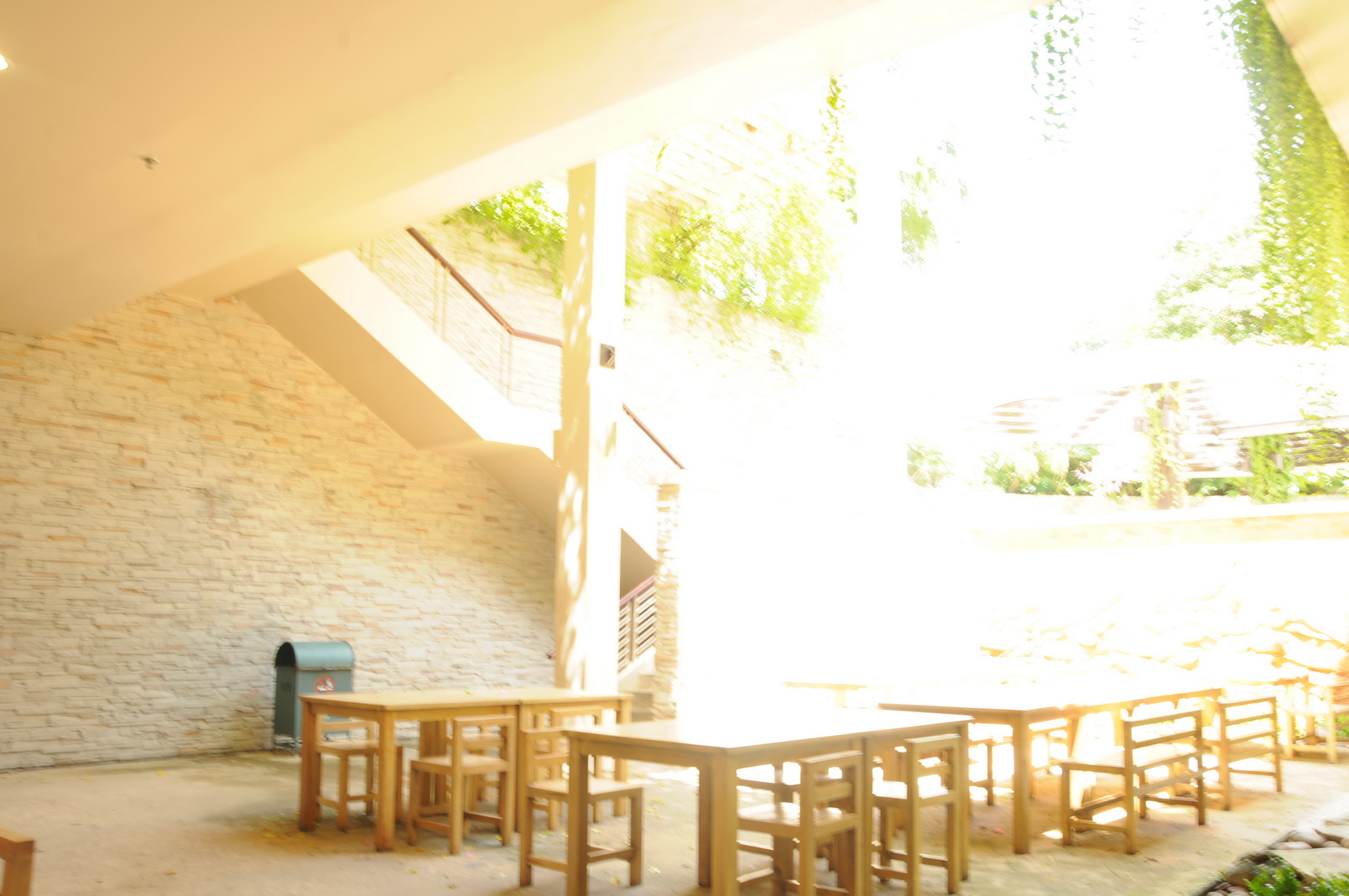} 		
		\includegraphics[width=0.108\linewidth]{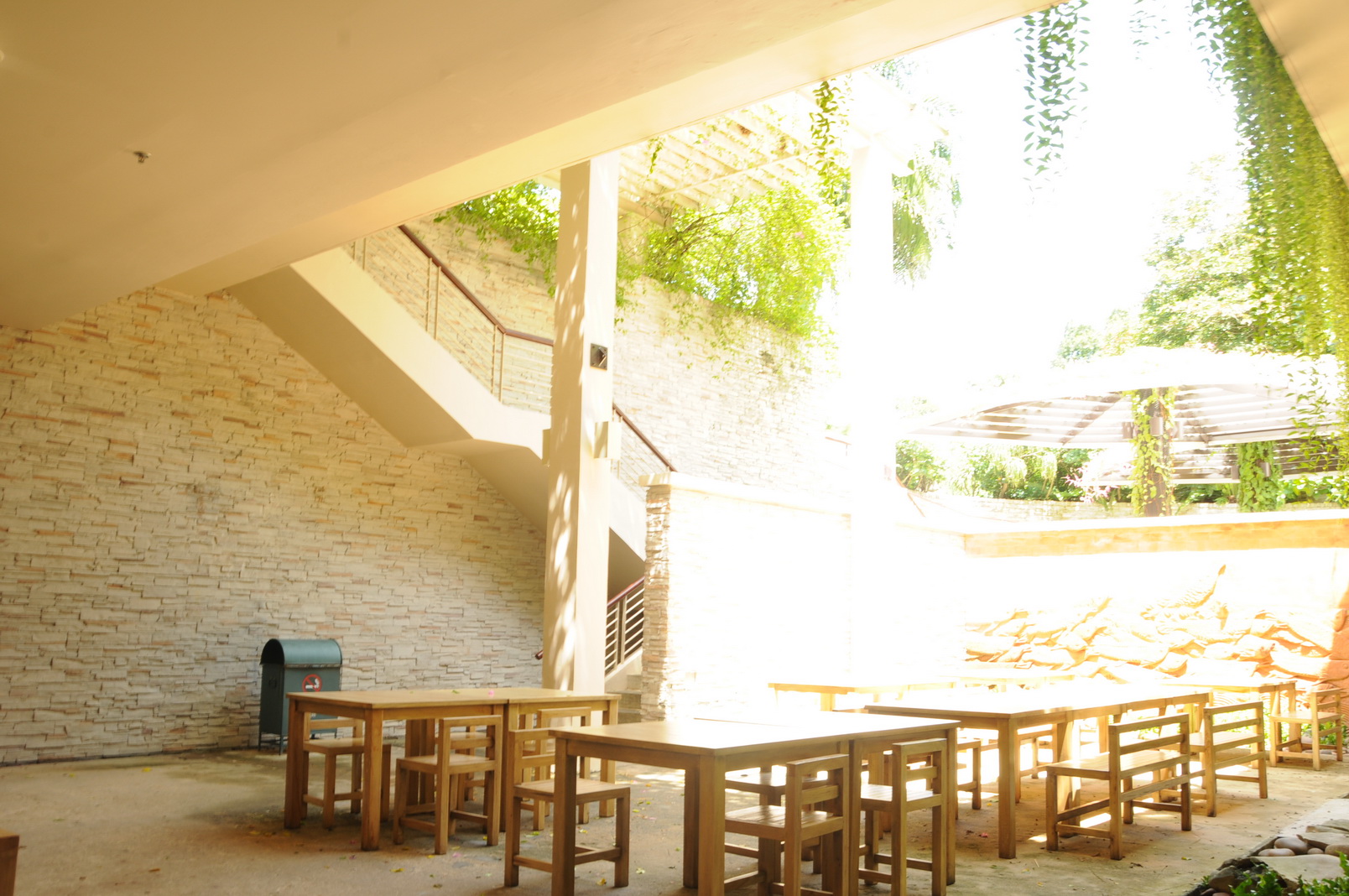}	
		\includegraphics[width=0.108\linewidth]{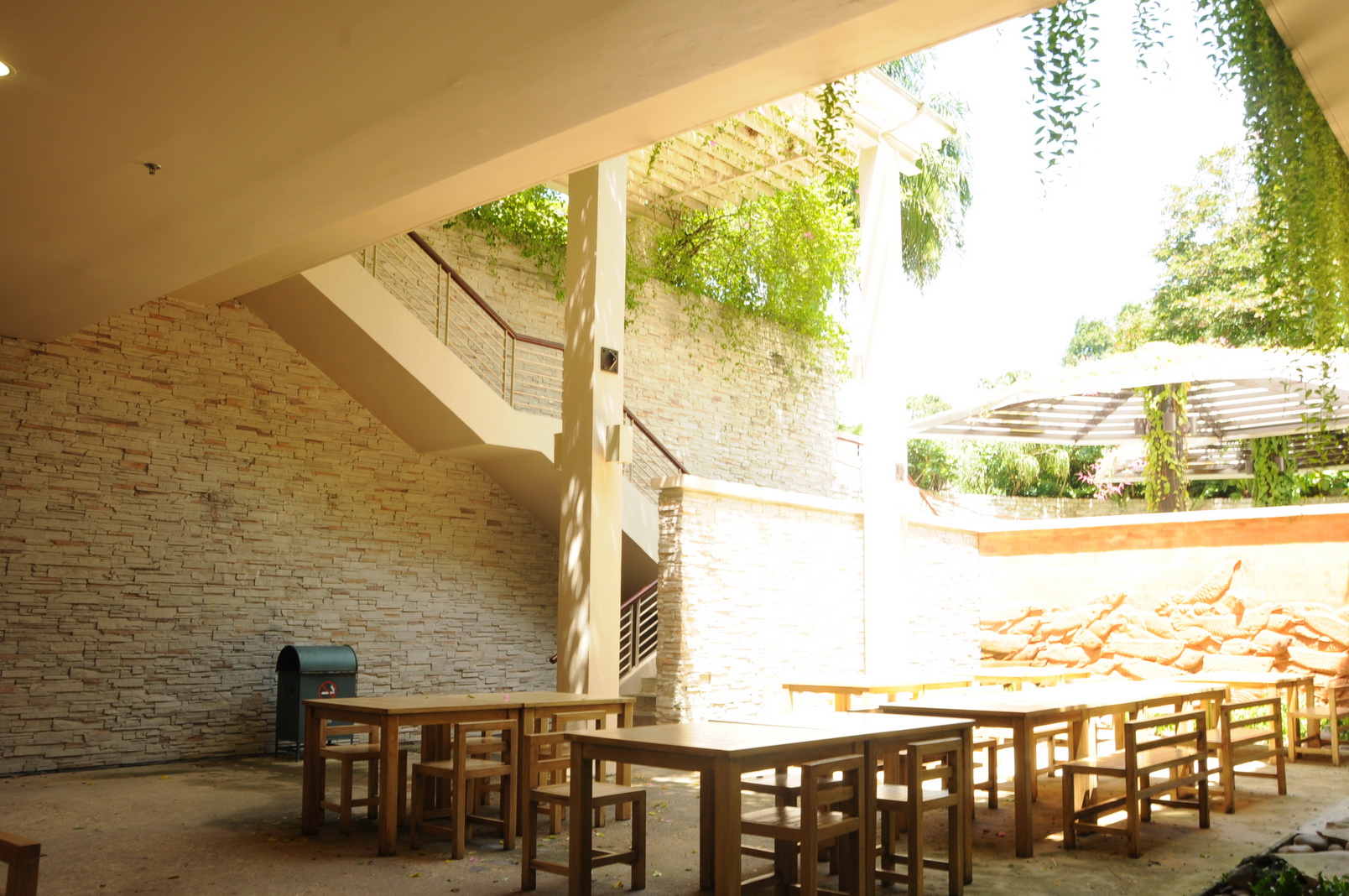}	
		\includegraphics[width=0.108\linewidth]{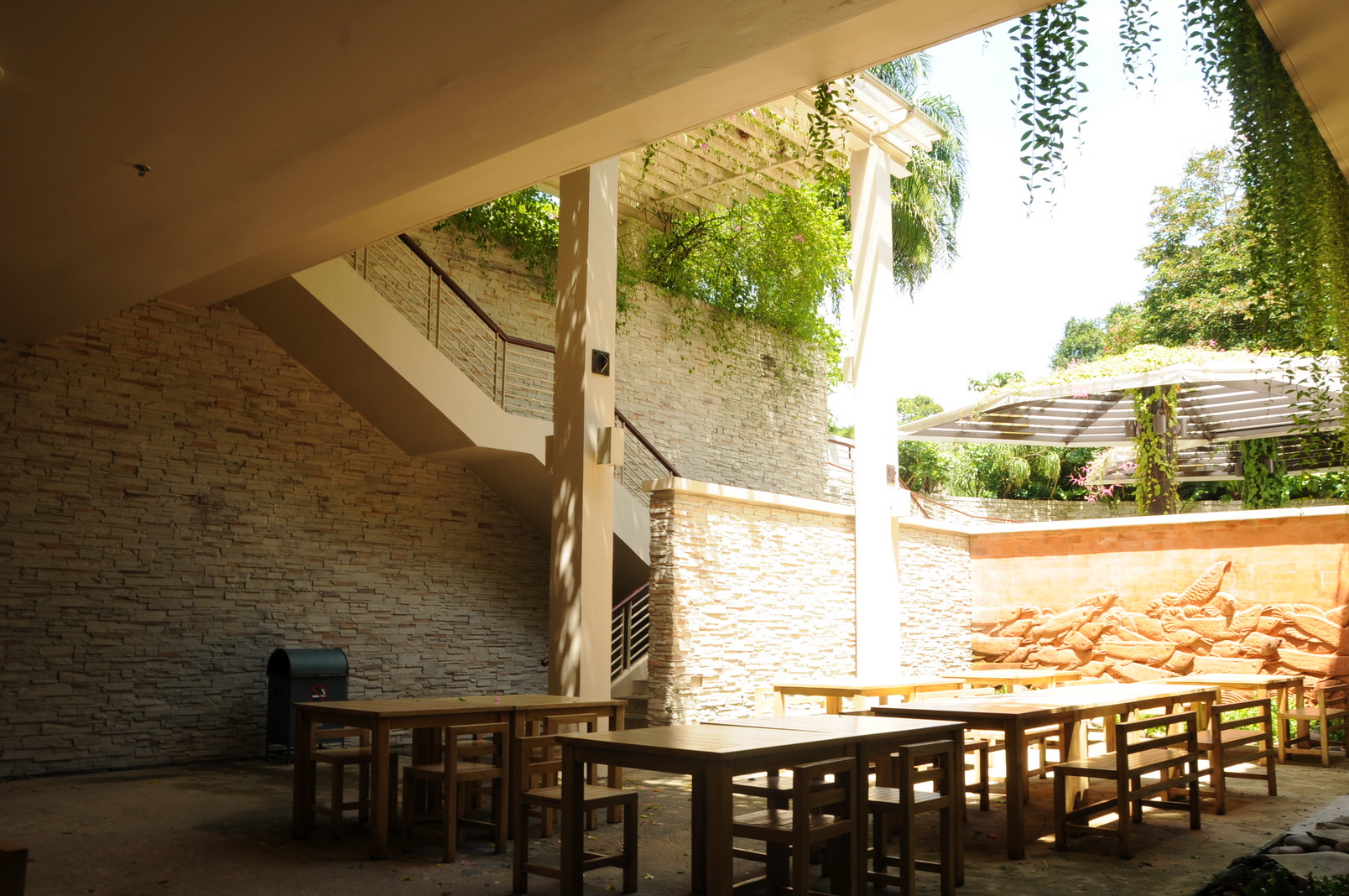}
		\includegraphics[width=0.108\linewidth]{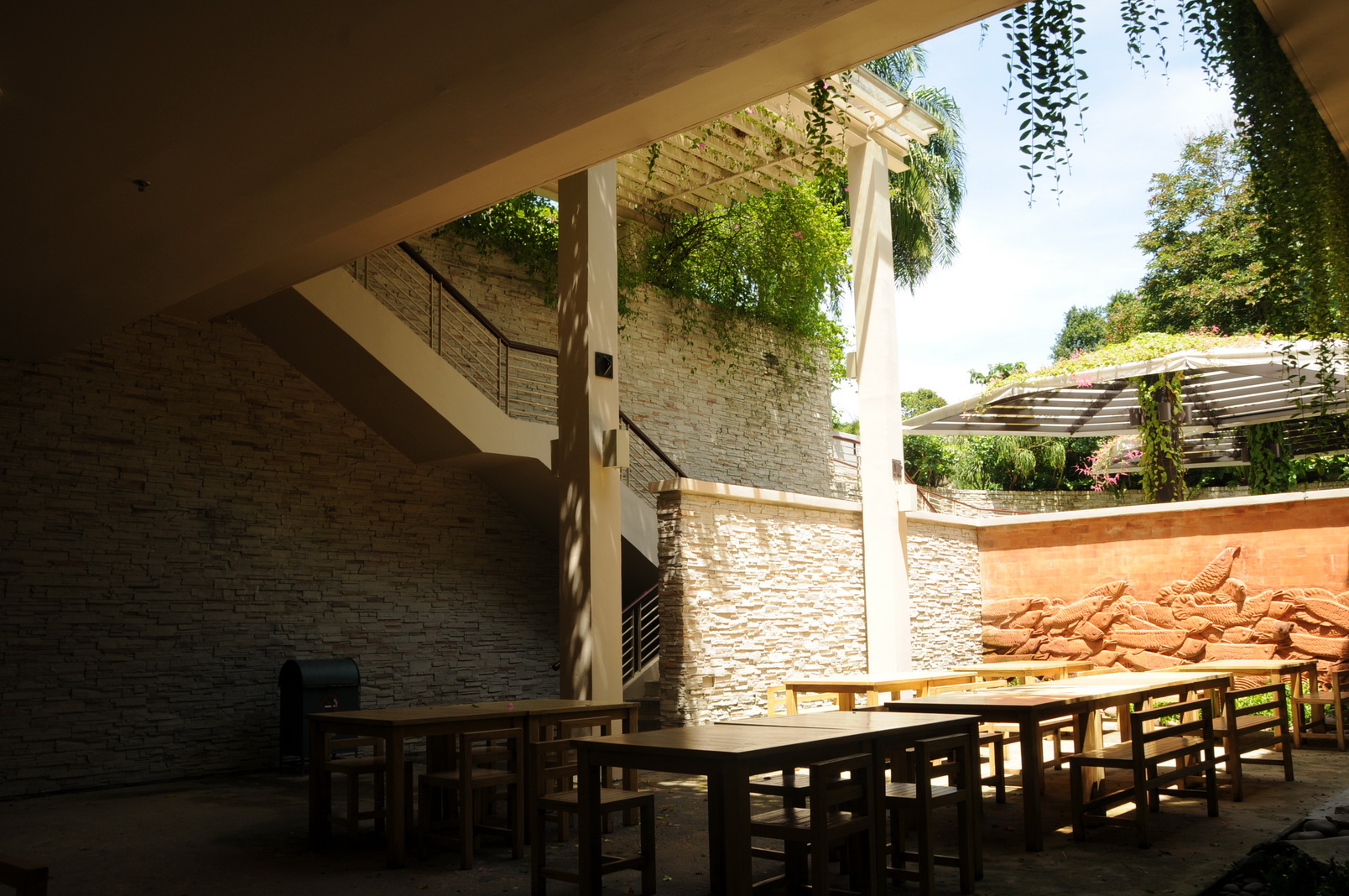}	
		\includegraphics[width=0.108\linewidth]{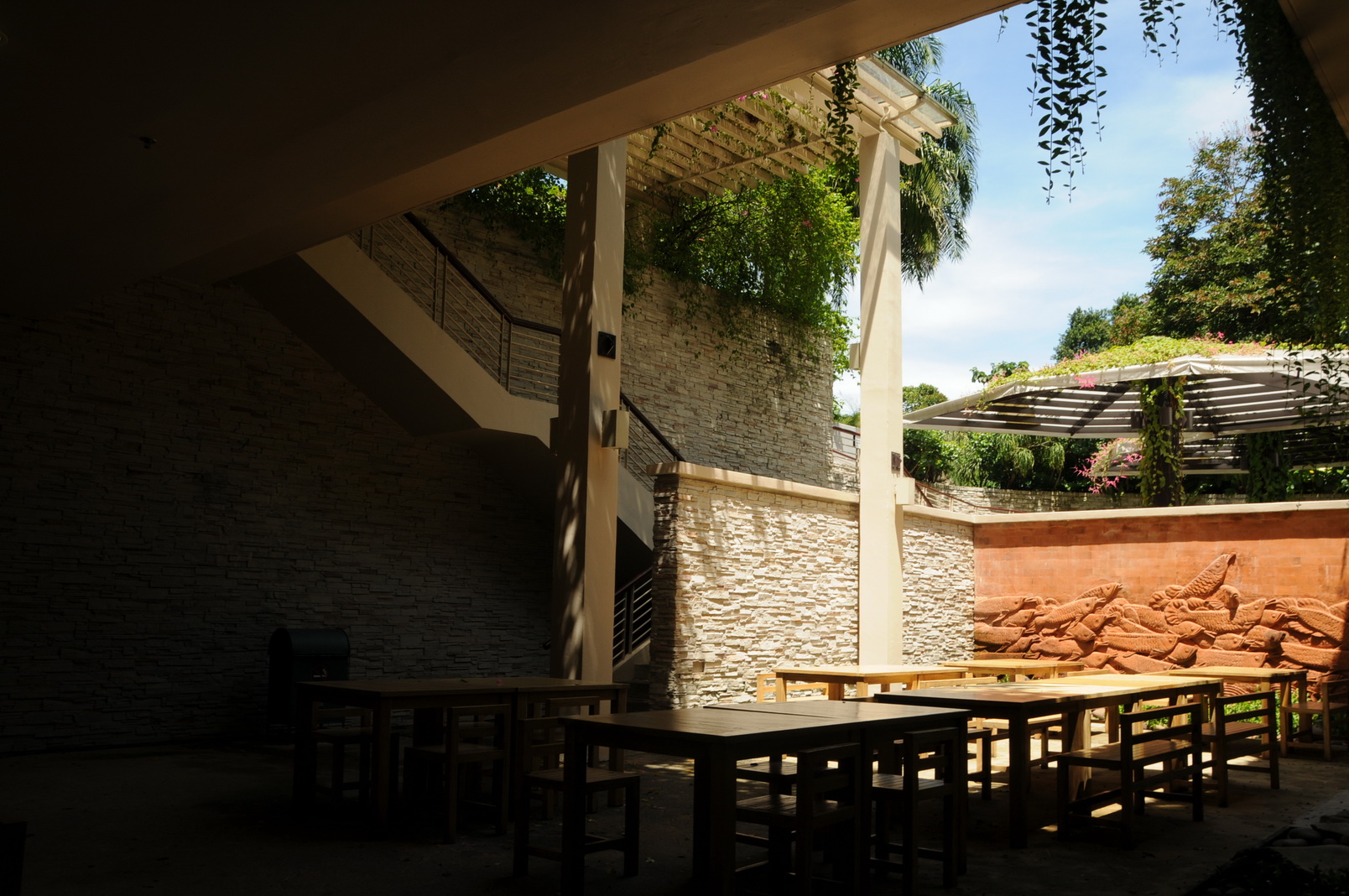}	
		\includegraphics[width=0.108\linewidth]{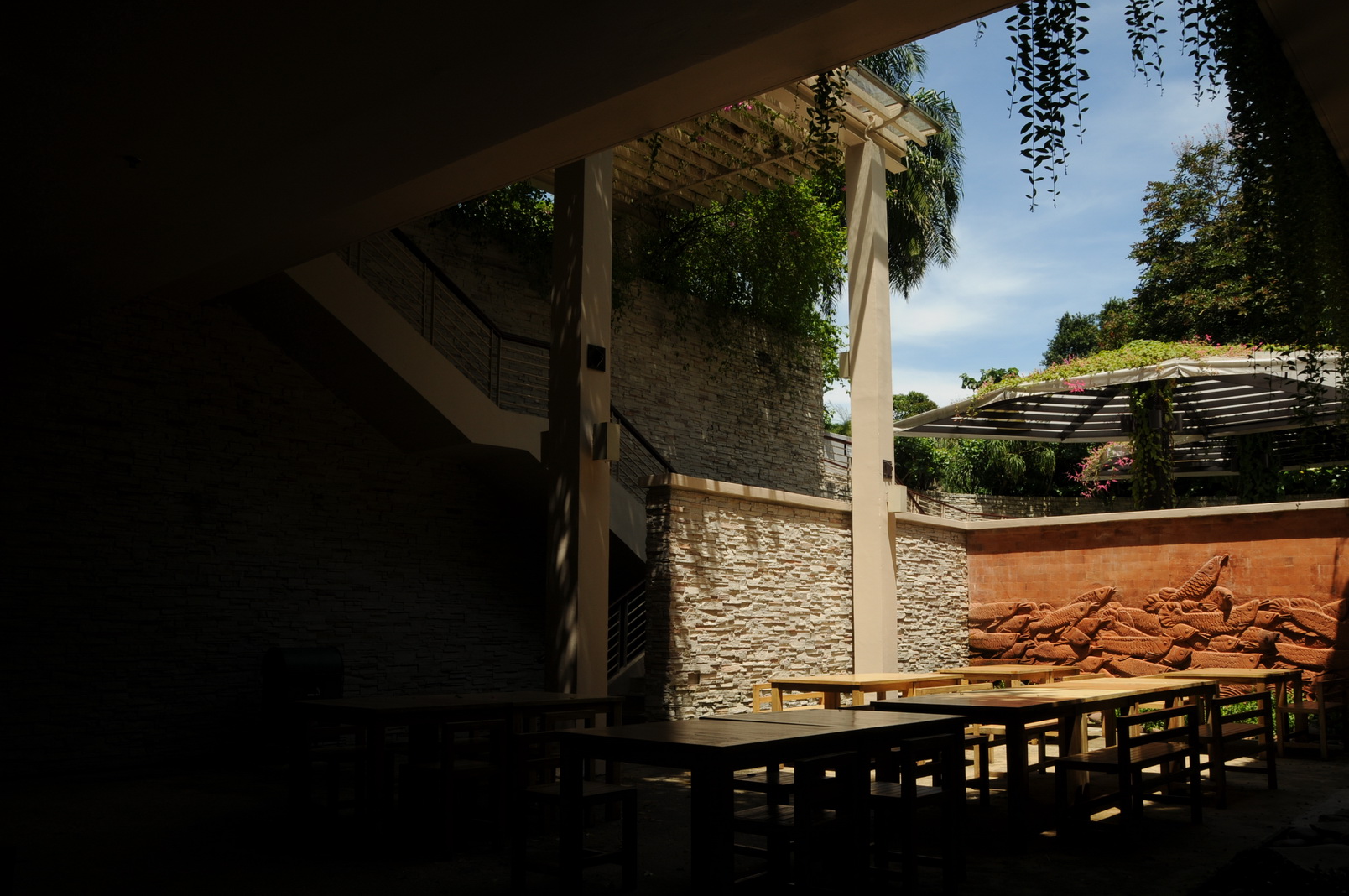}
		\includegraphics[width=0.108\linewidth]{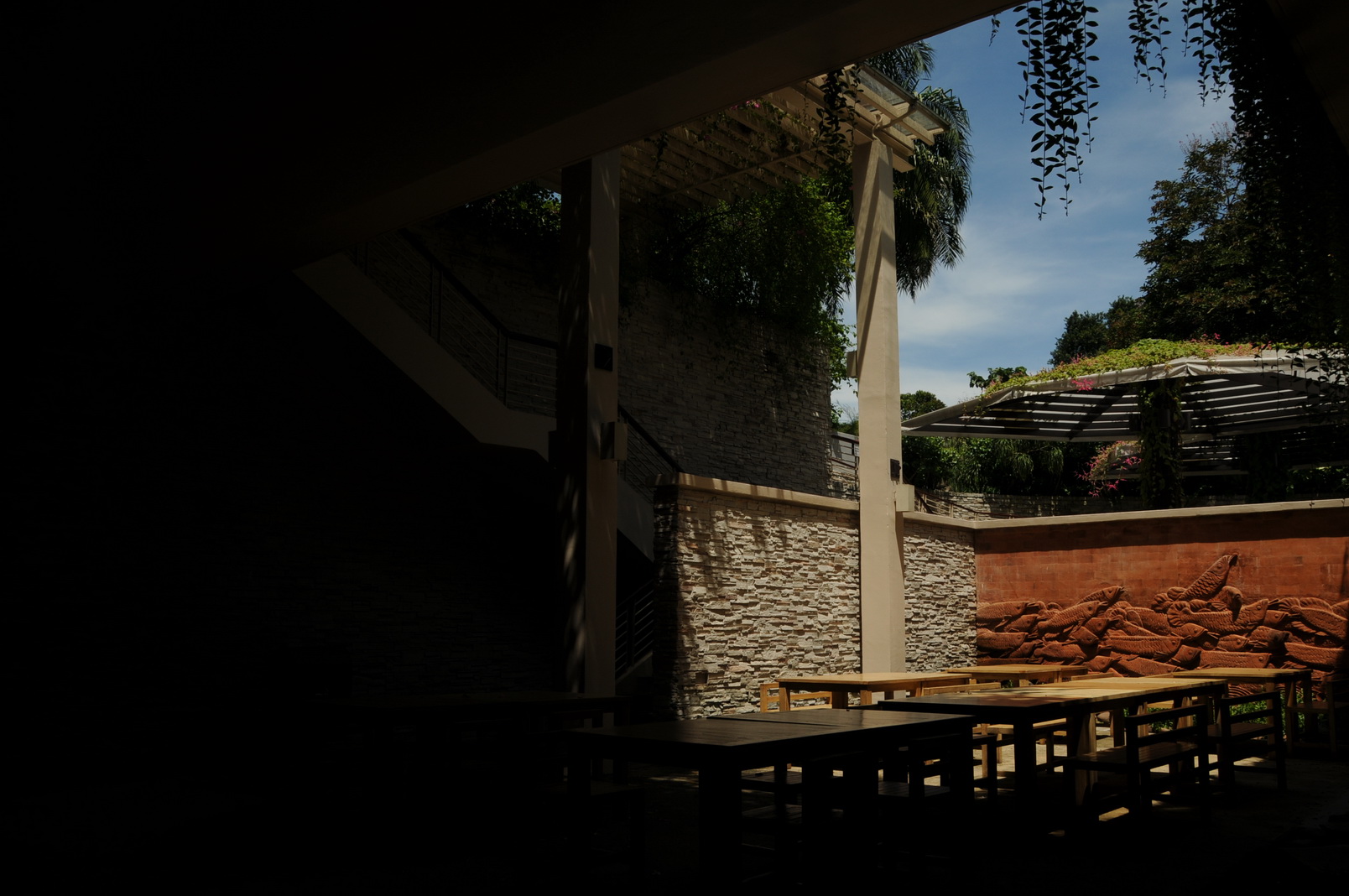}	
		\includegraphics[width=0.108\linewidth]{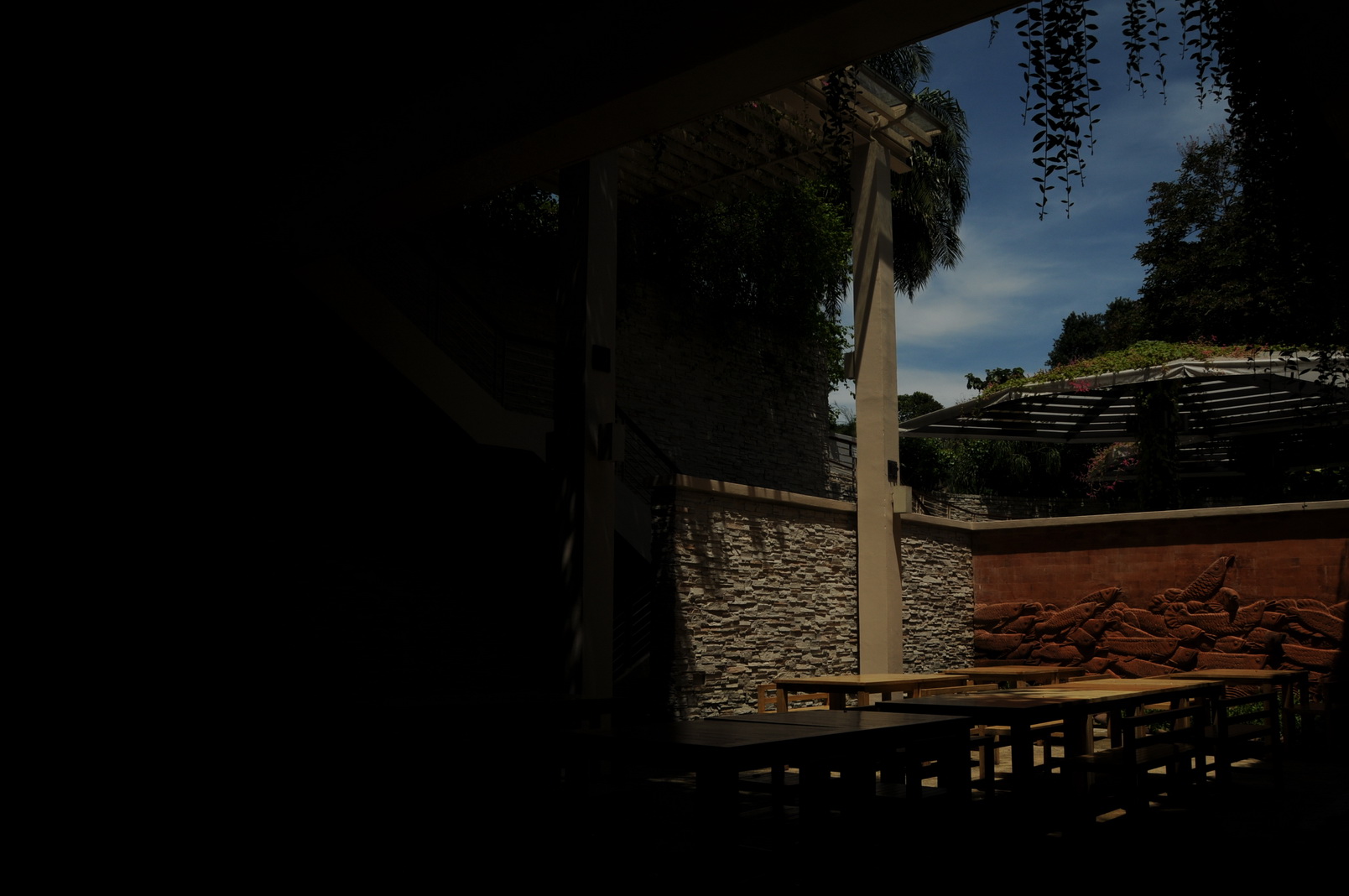}}
	
	\caption{Real multi-exposed images. Each sequence is arranged from long exposure to short exposure and the reference and first image in each sequence have over-exposed regions.}
	\label{Fig42}
\end{figure*}

\begin{figure*}[htb]
	\centering{
		\includegraphics[width=0.45\textwidth]{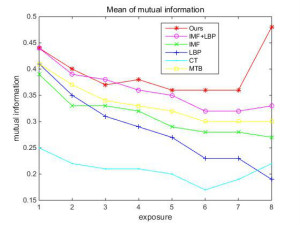}	
		\includegraphics[width=0.45\textwidth]{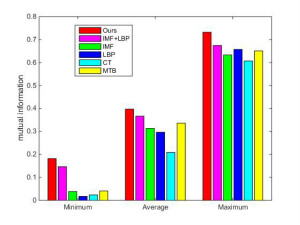}}	
	\caption{Mutual information for 35 real sequences.}
	\label{Fig2}	
\end{figure*}

For further objective evaluation, 35 sequences of real scenes are used here, which are captured using a NIKON D300, and a CANON EOS-1Ds Mark II, they are comprised of various scenarios including small/big camera motion, static/dynamic scenes, little/severe saturation, 1EV/2EV intervals and so on. Each sequence contains 4 to 11 images, two sample sequences of multi-exposure image datasets are presented in Fig. \ref{Fig42}. As motion parameters are unknown, alignment results are evaluated in terms of mutual information (MI) \cite{viola1997alignment}. Selecting the brightest image as the reference, the MIs in different exposures are shown in Fig. \ref{Fig2}. The performance of alignment methods usually decreases as the EV interval increases. This is because the normalization error increases with the increment of EV interval. On the other hand, the performances of the CT, IMF+LBP and the proposed method are robust to 7EV and 8 EV intervals, among which, the proposed method achieves the largest mutual information, especially when the EV interval between the two images is large.  The proposed normalization method reduces the errors of IMF in under/over exposed regions, and improves the similarity of two large-exposure-ratio images. It is also observed from Fig. \ref{Fig2} that the minimal, median and maximal MI of the proposed method is the largest, which implies the proposed method is suitable for various scene sets alignment and robust to intensity variations.

\subsection{tests on efficiency}

In order to demonstrate the efficiency of the proposed method, the operation times of different methods are calculated respectively. The learning-based methods are running in Python code\footnote{SuperPoint, https://github.com/magicleap/SuperPointPretrainedNetwork ;LF-Net, https://github.com/vcg-uvic/lf-net-release}, the rest are running in MATLAB code. The experiments are performed on DELL computer with i7-6700 CPU, 8GB RAM, and 64bit OS. However, the performances of MTB, LBP, CT, IMF, SIFT, BRIEF, IMF+BRIEF and the learning-based methods failed to align multi-exposed images in the previous experiment. The descriptor of IMF+BRIEF+``HD'' is a 256-vector which needs expensive computation to match two descriptors under ``Hamming distance''. Therefore, the operation times of the proposed method are compared with IMF+LBP and IMF+SIFT. The operation times (second) of different algorithms are tabulated in Table \ref{tab3}. 

Two sequences are randomly selected to test the efficiency. The results are given in Table \ref{tab3}. The operation time of the algorithm is proportional to image size and the number of features. IMF+SIFT is expensive in computation due to complex feature extraction and description as shown in Table \ref{tab4}. The normalization process of the proposed method is computationally more expensive than that of \cite{1wu2014}, but its alignment is 2.5 times faster than that of IMF+LBP and IMF+SIFT methods.

\begin{table}
	\centering
	\caption{Operation times (second) of different algorithms. All the codes are written in Matlab.}	
	\setlength{\tabcolsep}{3pt}
	\footnotesize
	\begin{tabular}{|c|c|c|c|c|c|}
		\hline		
		 Sequence       &No. of images  &Image size   & IMF+LBP & IMF+SIFT & Ours\\
		\hline
		 Snowman         &6  &693*505    &68.6    &183.2   &48.4\\
         Snow            &4  &960*680    &57.3    &1209  &36.6 \\	
		\hline 
	\end{tabular}
	\label{tab3}
\end{table}

\begin{table}
	\centering
	\caption{Operation times (second) of  normalization, feature extraction and description, alignment when aligning the first two ``Snowman'' images. All the codes are written in Matlab.}
	\setlength{\tabcolsep}{3pt}
	\footnotesize
	\begin{tabular}{|c|c|c|c|c|c|}
		\hline		
		Method       &Normalization  &Feature extraction and descripition  &Alignment\\
		\hline
		IMF+SIFT     &0.7  &59.8    &6.0 \\
		\hline
		IMF+LBP      &0.7  &\multicolumn{2}{c|} {16.3} \\
		Ours         &2.7  &\multicolumn{2}{c|} {6.1} \\		
		\hline 
	\end{tabular}
	\label{tab4}
\end{table}

\section{Conclusion}
\label{section4}

In this paper, a novel image alignment algorithm has been proposed for multi-exposed images with saturated regions by using a differentiable  ``Hamming'' distance.  The images are first normalized by using intensity mapping functions to reduce the effects of saturation. The normalized images are then coded by LBP and aligned by the proposed algorithm. Experimental results show that the proposed algorithm outperforms state-of-the-art alignment algorithms for differently exposed images. 

There are many applications of the proposed algorithm. For
example, it can be used in the autonomous SLAM \cite{1whel2015} with prior knowledge on the mapping environment, the direct visual odometry \cite{1alis2017} with the assistance of IMU integration \cite{John2018Motion}, and panorama imaging  for tele-operation of unmanned ground vehicles \cite{1yaow2015}. All these applications will be studied in our future research.

\end{document}